\documentclass[english,amssymb]{revtex4}

\usepackage{graphicx}
\usepackage{dcolumn}
\usepackage{bm}
\usepackage{physics}
\usepackage{xcolor}
\usepackage{mathtools}
\usepackage[colorlinks=true,linkcolor=blue]{hyperref}

\begin{document}

\title{Energy structure, density of states and transmission properties of \\ the periodic 1D Tight-Binding lattice with a generic unit cell of $u$ sites}

\author{K. Lambropoulos}
 \email{klambro@phys.uoa.gr}
\affiliation{National and Kapodistrian University of Athens, Department of Physics, Panepistimiopolis, 15784 Zografos, Athens, Greece}
\author{C. Simserides}
\email{csimseri@phys.uoa.gr}
\homepage{http://users.uoa.gr/~csimseri/physics_of_nanostructures_and_biomaterials.html}
\affiliation{National and Kapodistrian University of Athens, Department of Physics, Panepistimiopolis, 15784 Zografos, Athens, Greece}

\date{\today}

\begin{abstract}
We report on the electronic structure, density of states and transmission properties of the periodic one-dimensional Tight-Binding (TB) lattice with a single orbital per site and nearest-neighbor interactions, with a generic unit cell of $u$ sites. The determination of the eigenvalues is equivalent to the diagonalization of a real tridiagonal symmetric $u$-Toeplitz matrix with (cyclic boundaries) or without (fixed boundaries) perturbed upper right and lower left corners. We solve the TB equations via the Transfer Matrix Method, producing, analytical solutions and recursive relations for its eigenvalues, closely related to the Chebyshev polynomials. We examine the density of states and provide relevant analytical relations. We attach semi-infinite leads, determine and discuss the transmission coefficient at zero bias and investigate the peaks number and position, and the effect of the coupling strength and asymmetry as well as of the lead properties on the transmission profiles. We introduce a generic optimal coupling condition and demonstrate its physical meaning. 
\end{abstract}


\maketitle


\section{\label{sec:Introduction}Introduction}
Tight-Binding (TB) is a widely used method for determining the band structure through the expansion of the wavefunction in a basis of functions centered at each site. 
Decades after its foundation~\cite{SlaterKoster:1954}, it has been evolved into an efficient approach, employable to a broad class of problems regarding the electronic structure and properties of matter, requiring varying degrees of accuracy~\cite{Harrison,Papaconstantopoulos:2003}. TB gives realistic results when the spatial extent of the sites' wavefunctions is small compared to the distances between them, so that the overlap of the neighboring orbitals is small. TB is simple and computationally cheap, hence it can be applied to large systems. TB is widely used to describe, among others, polymers and organic systems. TB models are commonly applied to the study of charge transfer and transport properties of $\pi$-conjugated organic systems which are candidates for molecular or atomic wires, such as single and double stranded DNA chains~\cite{Roche-et-al:2003,Roche:2003,Shih:2008,Albuquerque:2014,Albuquerque:2005,Berlin:2002,Senthilkumar:2005,Cuniberti:2007,Simserides:2014,LKGS:2014,LChMKTS:2015, LKMTLGTCS:2016,LCMKLTTS:2016} as well as cumulenic and polyynic carbynes~\cite{Nikerov:1976,Tommasini:2007,Milani:2008,Tommasini:2008,La Magna:2009,Dyachkov:2013}. 

In the present work, we focus on the periodic 1D TB lattice with a generic unit cell composed of $u$ cites, one orbital per site and nearest neighbor interactions. Hence, we have $u$ different on-site energies and hopping integrals. We take explicitly into account the difference in the hopping integrals between different moieties inside the unit cell. The solution of the TB system of equations is equivalent to the diagonalization of a real symmetric tridiagonal $u$-Toeplitz matrix of order $N = mu$ with (cyclic boundaries) or without (fixed boundaries) perturbed upper right and lower left corners. 
The properties of such matrices have been extensively studied in the literature~\cite{Alvarez:2005,Yueh:2008,Kouachi:2006,Gover:1994,DaFonseca:2007} due to their theoretical interest and their many applications. We solve the TB system of equations via the transfer matrix method~\cite{Lee:1981,Stone:1981} and determine expressions for its eigenvalues, for both cyclic and fixed boundary conditions, by combining the spectral duality relations~\cite{Molinari:1997,Molinari:2003} with the connection of the elements of the powers of a $2\times2$ unimodular matrix to the Chebyshev polynomials of the second kind~\cite{Yeh:1977,Macia:2009}. For cyclic boundaries we derive the dispersion relation, while, for fixed boundaries we show that the spectrum of eigenvalues (eigenspectrum) is produced by a recurrent relation containing the Chebyshev polynomials of the second kind and the elements of the transfer matrix of the unit cell. We discuss the dispersion relations and eigenspectra through representative examples for systems with unit cells composed by $u=1,2,3,4$ sites, supposing different on-site energies and hopping integrals, that is, the most general case. We examine the density of states (DOS) and the occurrence of van Hove Singularities (VHS) and provide analytical relations for $u=1,2,3,4$. We determine the transmission coefficient (TC) at zero bias, by interconnecting the systems to semi-infinite leads. Many works have been dedicated to the TC determination, within various contexts cf. eg. Refs.~\cite{TsuandEsaki:1973,Emberly:1998,Kostyrko:2002,Smithetal:2002,Baringhausetal:2014,Macia:2005,GuoXu:2007a}. Here, we analyze in detail the transmission profiles of periodic TB systems, examining the influence of the strength and asymmetry of the system-leads coupling, the leads bandwidth and bandcenter, and the intrinsic interactions of the system. The number and position of the transmission peaks is determined in all cases. 
We introduce a generic optimal coupling condition and demonstrate its physical meaning. This condition generalizes for any periodic 1D TB chain the condition reported in Ref. \cite{Macia:2005}, where a periodic single-stranded DNA chain with $u=4$ and symmetric coupling to the leads was studied, with the hopping 
integrals of the system assumed identical.

The rest of the paper is organized as follows: 
in Sec. \ref{sec:TMM} we present the transfer matrix method and explain how the eigenvalues of the generic periodic system are determined, for cyclic and fixed boundaries. In Sec. \ref{sec:Evalues} we discuss dispersion relations and eigenspectra. 
Sec. \ref{sec:DOS} is devoted to the density of states. In Sec. \ref{sec:TC}, we determine the transmission coefficient and investigate all the factors that affect the transmission profiles. In Sec. \ref{sec:Conclusions}, we state our conclusions.

\section{\label{sec:TMM}The transfer matrix method}
The TB system of equations for a 1D lattice composed of $N$ sites with a single orbital per site and nearest neighbor interactions, also known as the Wire Model~\cite{Cuniberti:2007,LCMKLTTS:2016}, is 
\begin{equation} \label{Eq:TBsystem}
E \psi_n = \epsilon_n \psi_n + t_{n_-} \psi_{n-1} +  t_{n_+} \psi_{n+1},
\end{equation}
$ \forall n = 1, 2, \dots, N$, where $E$ is the eigenenergy, $\epsilon_n$ is the on-site energy at site $n$, 
$\psi_n = \braket{n}{\Psi}$, $\ket{\Psi}$ is the eigenfunction.
$|\psi_n|^2$ is the occupation probability at site $n$. 
$t_{n_+/n_-}$ is the hopping integral between sites $n$ and $n+1$ or $n$ and $n-1$, respectively. Let us set $\psi_{N+1} = \lambda \psi_{1}$ and $\psi_{0} = \lambda^* \psi_{N}$, where $\lambda = e^{ikNa}$ determines the boundary conditions. For cyclic boundaries, $k$ is a good quantum number and $\lambda = \lambda^* = 1$ (by Bloch theorem), while for fixed boundaries, $\lambda = \lambda^* = 0$. If the TB system is periodic, with a unit cell composed of $u$ sites, and is repeated $m$ times, so that $N = mu$, the solution of Eq. \eqref{Eq:TBsystem} is reduced to the diagonalization of the hamiltonian matrix 
\begin{equation}
H(\lambda) =
\begin{pmatrix}
\epsilon_1 & t_1 & & & & & & & \lambda^* t_u&\\
t_1 & \epsilon_1 & t_2 \\
& \ddots & \ddots &  \ddots\\
& & t_{u-2} & \epsilon_{u-1} & t_{u-1} \\
& & & t_{u-1} & \epsilon_u & t_u \\
& & & & t_u & \epsilon_1 & t_1 \\
& & & & & \ddots & \ddots & \ddots \\
& & & & & & \ddots & \ddots & \ddots \\
\lambda t_u & & & & & &  & t_{u-1} & \epsilon_u \\
\end{pmatrix},
\end{equation}
which is a tridiagonal real symmetric $u$-Toeplitz matrix of order $N$, with (for $\lambda = 1$, i.e. for cyclic boundaries) or without (for $\lambda = 0$, i.e. for fixed boundaries) perturbed upper left and lower right corners. Hence, the condition $\det(EI_N-H(\lambda)) = 0$, i.e. the roots of the characteristic polynomial of $H(\lambda)$, which is a polynomial of energy of degree $N$, gives the eigenvalues of the system.

Eq. \eqref{Eq:TBsystem} can equivalently be written in the form
\begin{equation} \label{Eq:TBsystemTM}
\begin{pmatrix}
\psi_{n+1}\\
\psi_n
\end{pmatrix}
=
\begin{pmatrix} 
\frac{E-\epsilon_n}{t_{n_+}} & -\frac{t_{n_-}}{t_{n_+}}\\
1 & 0
\end{pmatrix}
\begin{pmatrix}
\psi_n\\
\psi_{n-1}
\end{pmatrix} = Q_n(E) 
\begin{pmatrix}
\psi_n\\
\psi_{n-1}
\end{pmatrix},
\end{equation}
where $Q_n(E)$ is called the Transfer Matrix (TM) of site $n$. We can now iteratively define the Global Transfer Matrix (GTM), $\tilde{M}_N(E)$, of the periodic system as
\begin{equation} \label{Eq:TBsystemGTMdef}
\begin{pmatrix}
\psi_{N+1}\\
\psi_N
\end{pmatrix}
= \tilde{M}_N(E) 
\begin{pmatrix}
\psi_1\\
\psi_{0}
\end{pmatrix} \vcentcolon=
M_u(E)^m 
\begin{pmatrix}
\psi_1\\
\psi_{0}
\end{pmatrix}.
\end{equation}
$M_u(E)$ is the Unit Cell Transfer Matrix (UCTM)
\begin{equation} \label{Eq:TBsystemUCTMdef}
M_u(E) = \prod_{n=m}^{1}Q_n(E).
\end{equation}
Substituting $\psi_{N+1} = \lambda \psi_{1}$ and $\psi_{0} = \lambda^* \psi_{N}$, Eq. \eqref{Eq:TBsystemGTMdef} reads
\begin{equation} \label{Eq:TBsystemGTM}
M_u(E)^m 
\begin{pmatrix}
\psi_1\\
\psi_{0}
\end{pmatrix}
=\lambda
\begin{pmatrix}
\psi_{1}\\
\psi_0
\end{pmatrix}.
\end{equation}
The elements of the $2 \! \times \! 2$ UCTM are recurrently given by
\begin{subequations} \label{Eq:UCTMrec}
\begin{equation} \label{Eq:UCTMrec1}
M_u^{11(12)} = \frac{E-\epsilon_u}{t_u} M_{u-1}^{11(12)} -   \frac{t_{u-1}}{t_u} M_{u-2}^{11(12)} 
\end{equation}
\begin{equation} \label{Eq:UCTMrec2}
M_u^{21(22)} = M_{u-1}^{11(12)} 
\end{equation}
\end{subequations}
with initial conditions $M_1^{11} = (E-\epsilon_1)/t_1$, $M_1^{12} = -t_u/t_1$, $M_0^{21} = 1$, $M_0^{22} = 0$. Hence, $M_u^{11}$, $M_u^{12}$, $M_u^{21}$, and $M_u^{22}$, are polynomials of energy of degree $u$, $u-1$, $u-1$, and $u-2$, respectively. Since UCTM, and hence GTM, is a symplectic matrix, i.e. \begin{equation} \label{Eq:RUTMsymplectic}
M_u^T\begin{pmatrix}0 & 1\\-1 & 0\end{pmatrix}M_u = \begin{pmatrix}0 & 1\\-1 & 0\end{pmatrix},
\end{equation}
therefore UCTM, and hence GTM, is unimodular $(\det(M_u) = 1)$; if $w$ is an eigenvalue of $M_u$, so is $w^*$.~\cite{Molinari:1997,Molinari:2003}

\subsection{\label{subsec:CyclicBoundaries} Cyclic Boundaries}
For the cyclically bounded periodic system, Bloch theorem implies that $\lambda = 1 = e^{ikmua}$, where $ua$ is the lattice constant of the system. Hence, $k = \frac{2\pi \nu}{mua}$, where $\nu$ is an integer taking $m$ values, such that $k$ belongs to the $1^\text{st}$ Brillouin Zone (BZ). Thus, from Eq. \eqref{Eq:TBsystemGTM} it follows that
\begin{equation} \label{Eq:TBsystemRUTMpb}
M_u(E)
\begin{pmatrix}
\psi_1\\
\psi_{0}
\end{pmatrix}
= e^{ikua}
\begin{pmatrix}
\psi_{1}\\
\psi_0
\end{pmatrix}.
\end{equation}
From the symplectic property, we know that if $e^{ikua}$ is an eigenvalue of $M_u$, so is $e^{-ikua}$. Hence, the eigenvalues of $M_u$ are given by the condition
\begin{equation}
\det(e^{ikua}I_2 - M_u) = \det(e^{-ikua}I_2 - M_u) = 0,
\end{equation}
from which we can easily arrive at the dispersion relation
\begin{equation} \label{Eq:dispertionrelation}
\cos(kua) = \frac{\Tr(M_u)}{2} \vcentcolon= z(E).
\end{equation}
Substituting each of the $N$ energy eigenvalues obtained by the above dispersion relation to Eq.~\eqref{Eq:TBsystemTM}, leads, via its iterative application, to the construction of the full corresponding eigenvector of the hamiltonian matrix $H(\lambda = 1)$.~\cite{Molinari:2003} Hence, the solutions of Eq.~\eqref{Eq:dispertionrelation} produce the eigenvalues of the tridiagonal symmetric $u$-Toeplitz matrix of size $mu$ with perturbed corners. Additionally, from Eq.~\eqref{Eq:dispertionrelation} it follows that all eigenvalues are symmetric around the center of the $1^\text{st}$ BZ.

From the eigenvalues and eigenvectors of Eq.~\eqref{Eq:RUTMsymplectic}, it can also easily be shown (details e.g. in Ref.~\cite{Yeh:1977}) that the $m$-th power of the unimodular UCTM, $M_u(E)$, i.e. the GTM, $\tilde{M}_N(E)$, is given by~\cite{Macia:2009}
\begin{equation} \label{Eq:PowOfRUTM}
M_u^m = U_{m-1}(z) M_u - U_{m-2}(z) I,
\end{equation}
where $U_m(z)$ are the Chebyshev polynomials of the second kind of degree $m$  that satisfy the recurrence relation
\begin{equation} \label{Eq:ChebyshevUrec}
U_m(z)-2zU_{m-1}(z)+U_{m-2}(z) = 0, \forall m \ge 1,
\end{equation}
with initial conditions $U_1(z)=2z$, $U_0(z)=1$.~\cite{Mason:2003} 
Hence, the GTM of the cyclic system can be written as
\begin{equation} \label{Eq:GTMcyclicdef1}
\tilde{M}_N =
\begin{pmatrix}
U_{m-1}M_u^{11}-U_{m-2} & U_{m-1}M_u^{12} \\
U_{m-1}M_u^{21}			& U_{m-1}M_u^{22}-U_{m-2}
\end{pmatrix}
\end{equation}
or, alternatively, using Eq.~\eqref{Eq:ChebyshevUrec}, as
\begin{equation} \label{Eq:GTMcyclicdef2}
\tilde{M}_N =
\begin{pmatrix}
U_m - U_{m-1} M_u^{22} & U_{m-1} M_u^{12} \\
U_{m-1}M_u^{21}			& U_{m-1}M_u^{22}-U_{m-2}
\end{pmatrix}.
\end{equation}
Eq. \eqref{Eq:GTMcyclicdef2} and the formula connecting the Chebyshev polynomials of the first and second kind, $2T_m(z) = U_m(z)-U_{m-2}(z)$,~\cite{Mason:2003} lead to an alternate form from which the eigenvalues for cyclic boundaries can be found, i.e.
\begin{equation} \label{Eq:Eigenvaluespb}
T_m(z) = 1.
\end{equation}
The Chebyshev polynomials of the first kind are given by a recurrence relation identical to Eq.~\eqref{Eq:ChebyshevUrec}, with initial conditions $T_1(z)=z$,    $T_0(z)=1$.

\subsection{\label{subsec:FixedBoundaries}Fixed Boundaries}
For the system with fixed boundaries $\lambda=0$. Therefore, Eq. \eqref{Eq:TBsystemGTM} reads
\begin{equation} \label{Eq:TBsystemRUTMfb}
\tilde{M}_N
\begin{pmatrix}
\psi_1\\
0
\end{pmatrix}
= 0
\begin{pmatrix}
\psi_{1}\\
0
\end{pmatrix} \Rightarrow \tilde{M}_N^{11} \psi_1 = 0 \psi_1. 
\end{equation}
Hence, $\psi_1$ is an element of the eigenvector of $H(\lambda = 0)$ if the eigenvalues $E$ of $H(\lambda = 0)$ satisfy the condition $\tilde{M}_N^{11}(E) = 0$. Since both $\tilde{M}_N^{11} = 0$ and $\det(EI_N-H(0)) = 0$ are polynomial equations of $E$ of the same degree ($N$) that are mutually satisfied \cite{Molinari:1997}, Eq. \eqref{Eq:TBsystemRUTMfb} produces the eigenvalues of the tridiagonal symmetric $u$-Toeplitz matrix of size $mu$. Exploiting Eqs.~\eqref{Eq:GTMcyclicdef1}-\eqref{Eq:GTMcyclicdef2}, the determination of these eigenvalues is reduced to the solution of
\begin{equation} \label{Eq:Eigenvaluesfb1}
U_{m-1}(z) M_u^{11} = U_{m-2}(z),
\end{equation}
or, alternatively
\begin{equation} \label{Eq:Eigenvaluesfb2}
U_m(z) = U_{m-1}(z) M_u^{22}.
\end{equation}

\subsection{\label{subsec:phonons} Phonons}
Closing this Section, we should notice that the above analysis holds also for the problem of phonon dispersion relations in periodic 1D Lattices with nearest neighbor interactions. The only thing that changes is the form of the TM of the sites, i.e. $Q_n$ in Eq. \eqref{Eq:TBsystemTM} takes the form
\begin{equation}
Q_n(\omega_{\textrm{ph}}) = \begin{pmatrix}
1 + \frac{K_{n_-}}{K_{n_+}} \! - \! \frac{M_n \omega_{\textrm{ph}}^2}{K_{n_+}} & -\frac{K_{n_-}}{K_{n_+}}\\
1 & 0
\end{pmatrix}.
\end{equation}
$\omega_{\textrm{ph}}$ is the phonon frequency, $M_n$ is the mass at site $n$ and $K_{n_\pm}$ are the spring constants connecting nearest neighboring sites and $\psi_n$ then correspond to the oscillation amplitudes.

\section{\label{sec:Evalues} Dispersion relations and Eigenspectra}
As shown in Sec. \ref{sec:TMM}, the eigenvalues of any periodic 1D lattice with a single orbital per site and nearest-neighbor interactions, for either cyclic or fixed boundary conditions, can be determined by Eqs. \eqref{Eq:dispertionrelation} or \eqref{Eq:Eigenvaluesfb1}-\eqref{Eq:Eigenvaluesfb2}, respectively, via which, the problem is reduced to the determination of the elements of UCTM using Eqs. \eqref{Eq:UCTMrec}. We have verified that the eigenvalues obtained by the above mentioned equations coincide with those obtained by the numerical diagonalization of matrix $H(\lambda=0,1)$. In Appendix \ref{ApA}, we provide the general form of the UCTM for systems with unit cell composed of $u = 1,2,3,4$ sites [Eqs. \eqref{Eq:UCTM1}-\eqref{Eq:UCTM4}], with the difference in the hopping integrals between differing sites inside the unit cell being explicitly taken into account. Below, we discuss on the eigenvalues of such systems, as the simplest examples of the generic periodic system.

\subsection{\label{subsec:Evalues1} Unit cell of one site ($u=1$, $m=N$)}
\begin{figure} [h!]
\centering
\includegraphics[width=0.9\columnwidth]{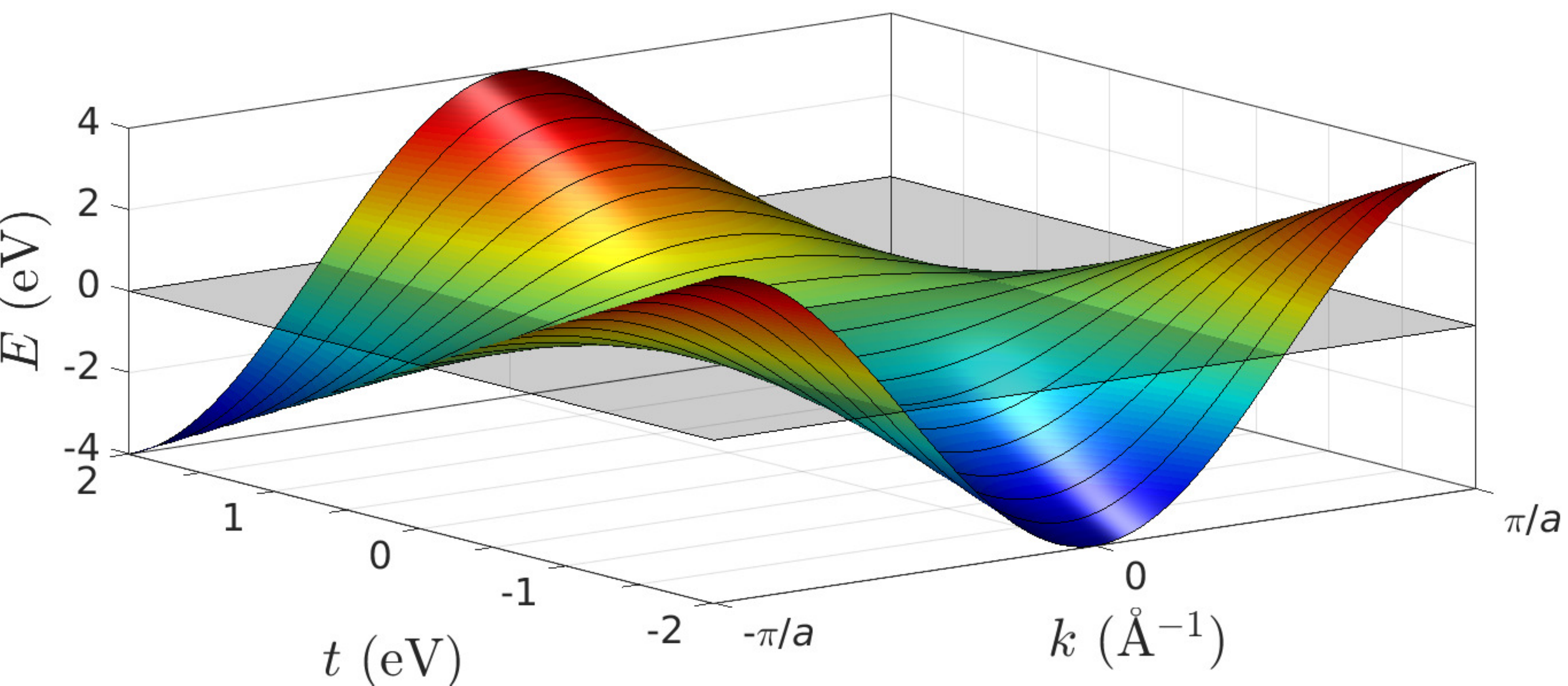}
\caption{Band structure of the 1D periodic lattice 
with \textit{cyclic boundaries} (Eq. \eqref{Eq:evsUCTM1pb}), 
with a unit cell of one site ($u = 1$), 
for a fixed on-site energy $\epsilon_1 = 0$, 
as a function of the hopping integral $t_1$.}
\label{fig:typadisprel3d}
\end{figure}

For \textit{cyclic boundaries}, the dispersion relation of Eq. \eqref{Eq:dispertionrelation} takes the well known form
\begin{equation} \label{Eq:evsUCTM1pb}
E = \epsilon_1 + 2t_1\cos(ka) 
  = \epsilon_1 + 2t_1\cos(\frac{2\pi\nu}{N}).
\end{equation}
$\nu=1,2,\dots,N$.
The dispersion relation is presented in Fig. \ref{fig:typadisprel3d}, for a fixed on-site energy, $\epsilon_1 = 0$, where the effect of the hopping integral $t_1$ alteration is also demonstrated. Increasing $\abs{t_1}$ linearly increases the bandwidth. Although changing the sign of $t_1$ does not affect the eigenenergies' values, it moves the position of their extrema from the middle to the edge of the $1^\text{st}$ BZ.

\begin{figure} [h!]
\centering
\includegraphics[width=0.9\columnwidth]{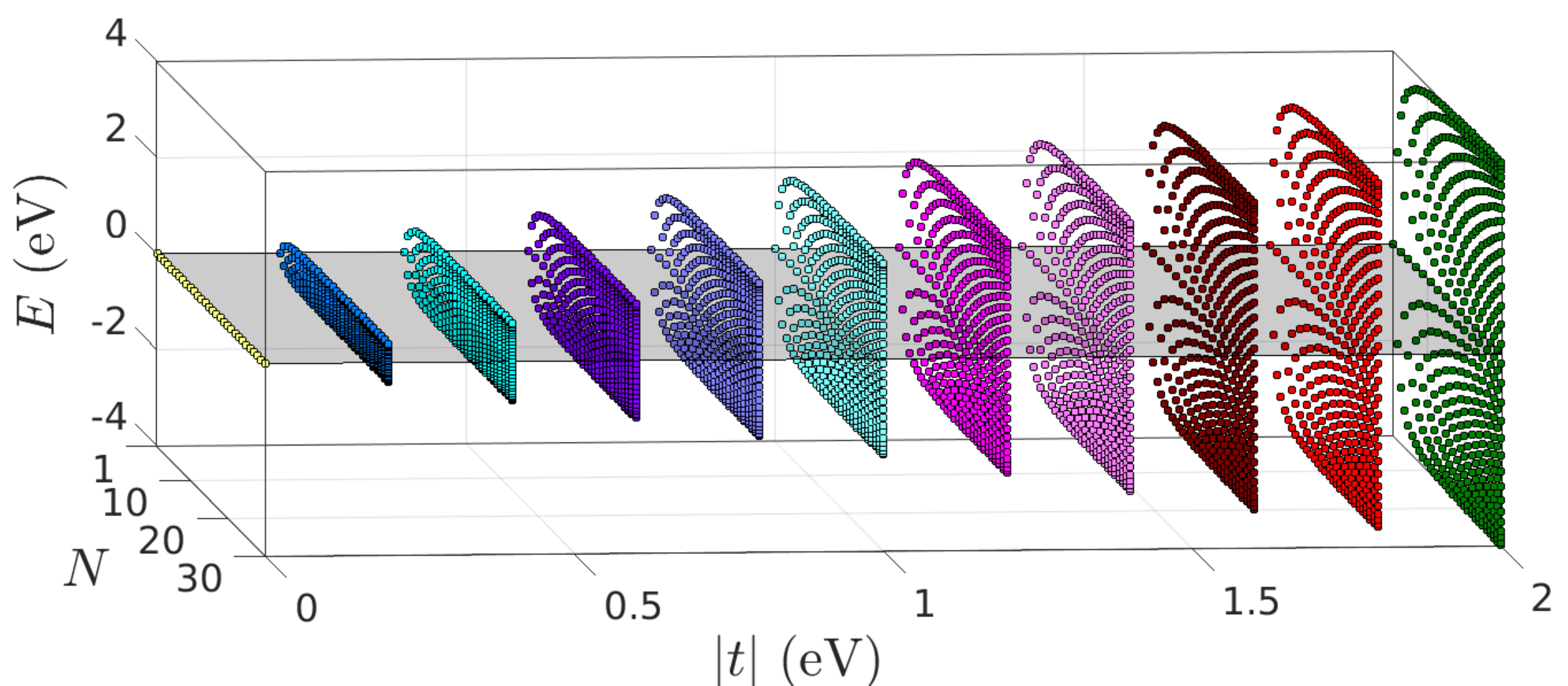}
\caption{Eigenspectrum of the 1D periodic lattice 
with \textit{fixed boundaries} (Eq. \eqref{Eq:evsUCTM1fb}),
with a unit cell of one site ($u = 1$), for $N$ up to 30 sites, 
for a fixed on-site energy $\epsilon_1 = 0$, 
as a function of the absolute value of the hopping integral $\abs{t_1}$.}
\label{fig:typeaeigenspec3d}
\end{figure}

For \textit{fixed boundaries}, it follows from Eq. \eqref{Eq:Eigenvaluesfb2} that the eigenvalues are produced by the zeros of $U_N(z)$, i.e. $z(E) = \cos(\frac{\mu\pi}{N+1}), \forall \mu=1,2,\dots,N$,~\cite{Mason:2003} hence
\begin{equation} \label{Eq:evsUCTM1fb}
E= \epsilon_1 + 2t_1\cos(\frac{\mu\pi}{N+1}).
\end{equation}
The eigenspectrum for fixed boundaries as $N$ increases, is presented in Fig. \ref{fig:typeaeigenspec3d}, for a fixed on-site energy, $\epsilon_1 = 0$, where the effect of the hopping integral absolute value $\abs{t_1}$ alteration is also demonstrated.

Eqs.~\eqref{Eq:evsUCTM1pb} and \eqref{Eq:evsUCTM1fb} can be derived analytically by direct diagonalization of the hamiltonian matrices cf. e.g. ~Ref.~\cite{LChMKTS:2015}. We have just demonstrated how they are derived with the transfer matrix method.


\subsection{\label{subsec:Evalues2} Unit cell of two sites ($u=2$, $m={\frac{N}{2}}$)}
\begin{figure} [h!]
\centering
\includegraphics[width=0.9\columnwidth]{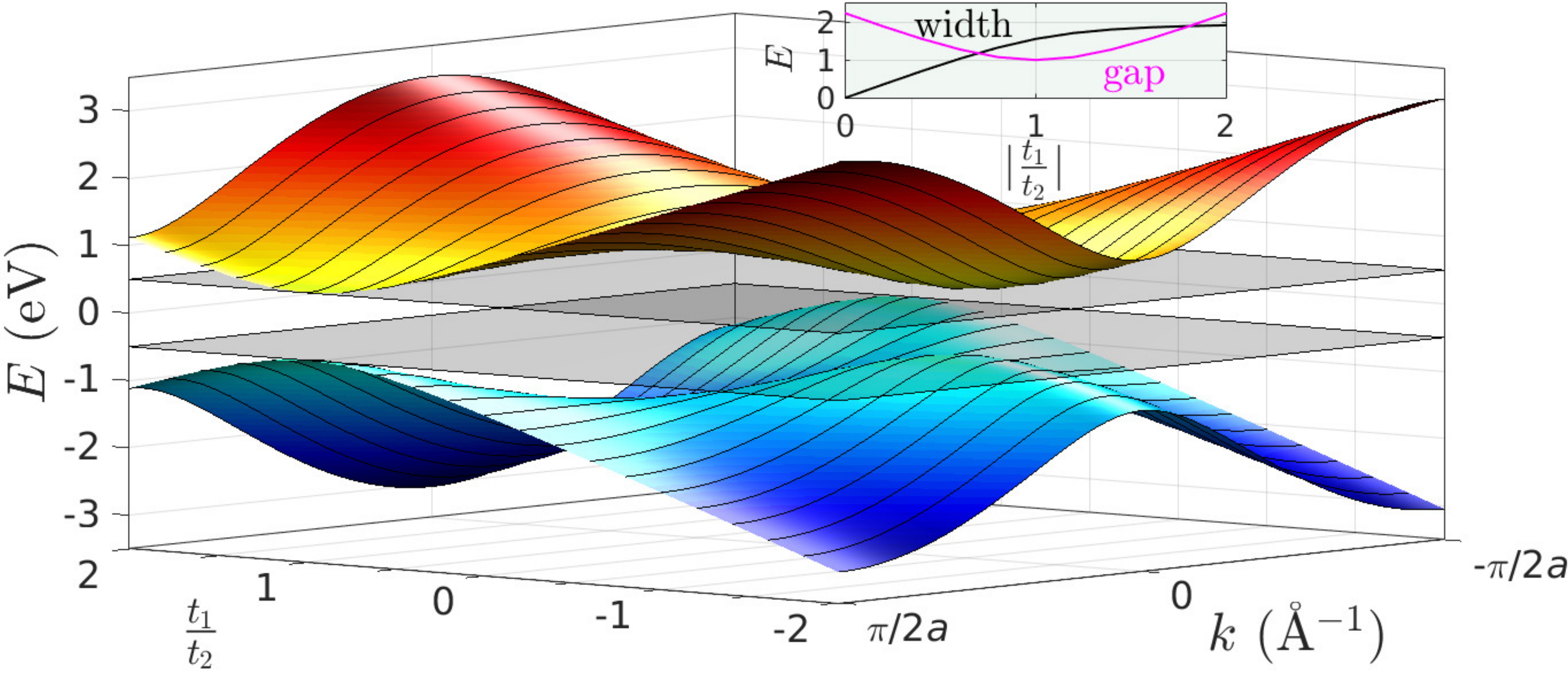}
\caption{Band structure of the 1D periodic lattice with a unit cell of two sites ($u=2$) with \textit{cyclic boundaries}   
(Eq. \eqref{Eq:evsUCTM2pb}), 
for fixed on-site energies $(\epsilon_1,\epsilon_2) = (-0.5,0.5)$ eV, 
as a function of the ratio between the hopping integrals $\frac{t_1}{t_2}$. 
(Inset) The bandwidths and the bandgap as a function of $\abs{t_1/t_2}$.}  
\label{fig:typecdisprel3d}
\end{figure}

For \textit{cyclic boundaries}, Eq. \eqref{Eq:dispertionrelation} takes the form
\begin{equation} \label{Eq:evsUCTM2pb}
E = \frac{\epsilon_1+\epsilon_2}{2} \pm \sqrt{\left(\frac{\epsilon_1-\epsilon_2}{2}\right)^2 + t_1^2 +t_2^2+2t_1t_2\cos(k2a)}.
\end{equation}
This dispersion relation is presented in Fig. \ref{fig:typecdisprel3d}, for fixed on-site energies $(\epsilon_1,\epsilon_2) = (-0.5,0.5)$ eV, where the effect of altering the ratio between the hopping integrals $\frac{t_1}{t_2}$, is also demonstrated. Again, changing the sign of $\frac{t_1}{t_2}$ moves the position of the eigenvalues' extrema from the middle to the edge of the $1^\text{st}$ BZ. As $\abs{\frac{t_1}{t_2}}$ increases from $0$ to $1$ the widths of the two bands (which are equal to each other) increase and the bandgap decreases until it gets equal to the energy gap between the on-site energies (for $\abs{\frac{t_1}{t_2}} = 1$). For $\abs{\frac{t_1}{t_2}} >1$ the bandwidths continue to increase, albeit much slower, and the bandgap starts to increase as well.

\begin{figure} [h!]
\centering
\includegraphics[width=0.9\columnwidth]{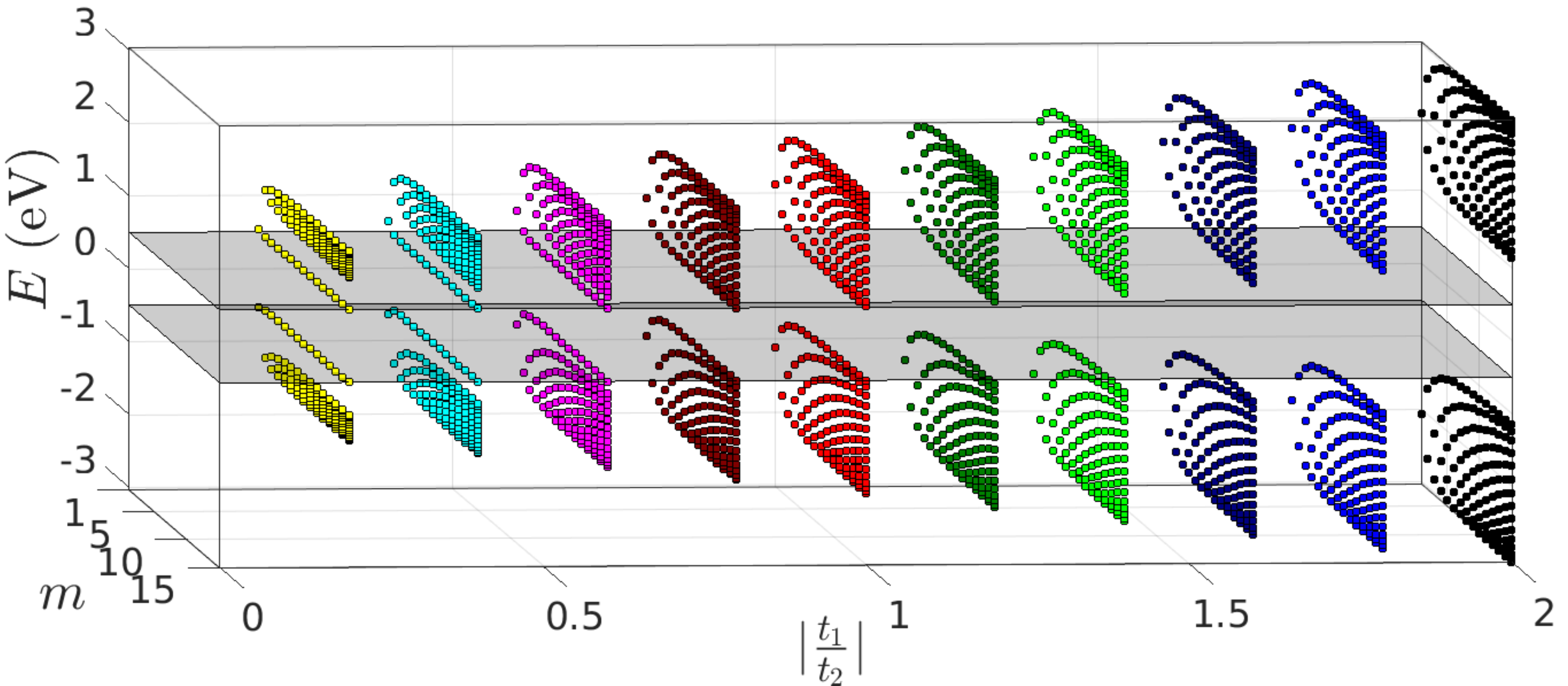}
\caption{Eigenspectrum of the 1D periodic lattice with a unit cell of two sites ($u=2$) with \textit{fixed boundaries} (Eq.\eqref{Eq:evsUCTM2fb}),  
for $m$ up to 15 unit cells ($N = 30$), 
for fixed on-site energies $(\epsilon_1,\epsilon_2) = (-0.5,0.5)$ eV, as a function of the absolute ratio between the hopping integrals $\abs{t_1/t_2}$.} 
\label{fig:typeceigenspec3d}
\end{figure}

For \textit{fixed boundaries}, it follows from Eq. \eqref{Eq:Eigenvaluesfb2} that the eigenvalues are the solutions of
\begin{equation} \label{Eq:evsUCTM2fb}
U_{m-1}(x) = -\frac{t_1}{t_2}U_m(x).
\end{equation}
The eigenspectrum for $m$ up to 15 unit cells ($N$=30) is shown in Fig. \ref{fig:typeceigenspec3d}, for fixed on-site energies $(\epsilon_1,\epsilon_2) = (-0.5,0.5)$ eV, where the effect of the absolute ratio between the hopping integrals $\abs{\frac{t_1}{t_2}}$ alteration is also demonstrated. From Fig. \ref{fig:typeceigenspec3d} we notice that as the value of $\abs{\frac{t_1}{t_2}}$ decreases down to 1, the bandgap decreases until it gets equal, for large $N$, to the energy gap between the on-site energies (for $\abs{\frac{t_1}{t_2}} = 1$). Further decrease of $\abs{\frac{t_1}{t_2}}$ below $1$ leads to an increase in the gap between the main bands, but there is a single additional eigenvalue in the upper (lower) band that, for large $N$, reaches the value $\epsilon_1$ ($\epsilon_2$). This behavior does not occur for cyclic boundaries, hence these states inside the band gap arise solely from the boundary effect. As it will be shown below, this effect occurs generally for fixed boundaries, although it arises from more complex relations between the parameters of the system.



\subsection{\label{subsec:Evalues3} Unit cell of three sites ($u=3$, $m={\frac{N}{3}}$)}
\begin{figure} [h!]
\centering
\includegraphics[width=0.9\columnwidth]{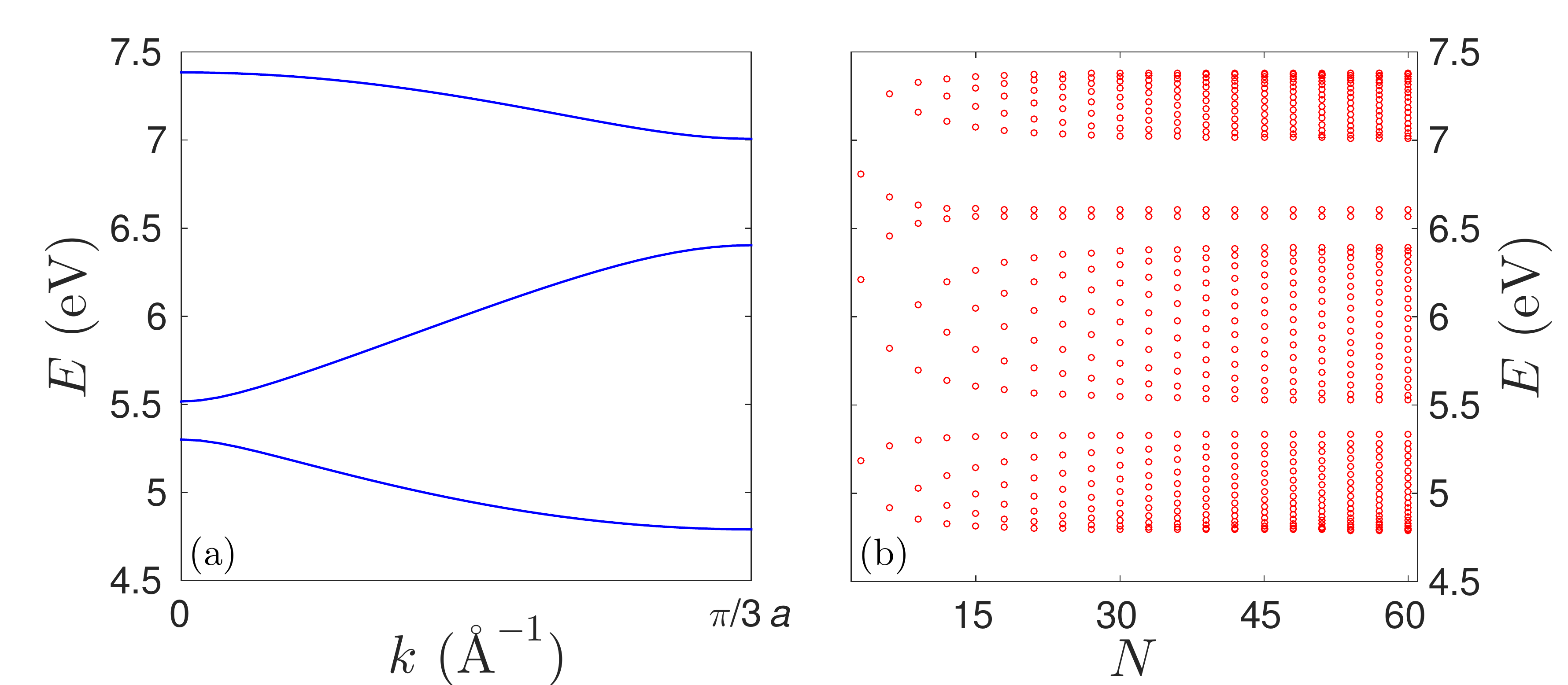}
\caption{(a) Dispersion relation for \textit{cyclic boundaries} and (b) eigenspectrum for \textit{fixed boundaries} for $m$ up to 20 unit cells ($N = 60$) of a 1D periodic lattice with a unit cell of three sites, with on-site energies $(\epsilon_1,\epsilon_2,\epsilon_3) = (6.3,5.8,6.1)$ eV and hopping integrals $(t_1,t_2,t_3) = (0.5,0.6,0.8)$ eV, as determined by Eqs. \eqref{Eq:evsUCTM3pb} and \eqref{Eq:evsUCTM3fb}, respectively.} 
\label{fig:UC3eigenvalues}
\end{figure}

For \textit{cyclic boundaries}, Eq. \eqref{Eq:dispertionrelation} takes the form
\begin{equation} \label{Eq:evsUCTM3pb}
\cos(k3a) = \frac{1}{2} \prod_{i=1}^{3}\frac{(E-\epsilon_i)}{t_i} -\sum_{ijk}^{\circlearrowright^3}\frac{(E-\epsilon_i)t_j}{2t_kt_i},
\end{equation}
where $\circlearrowright^3$ denotes all cyclic permutations of the indices $i \neq j \neq k \in \{1,2,3\}$. This dispersion relation is presented in Fig \ref{fig:UC3eigenvalues}(a), for the example of a system with on-site energies $(\epsilon_1,\epsilon_2,\epsilon_3) = (6.3,5.8,6.1)$ eV and hopping integrals $(t_1,t_2,t_3) = (0.5,0.6,0.8)$ eV. The three bands that occur for this system have widths $\approx 0.51, 0.89,0.38$ eV (from lower to higher energy) and the corresponding gaps are $\approx 0.22, 0.60$ eV.

For \textit{fixed boundaries}, it follows from Eq. \eqref{Eq:Eigenvaluesfb2} that the eigenvalues are the solutions of
\begin{equation} \label{Eq:evsUCTM3fb}
U_m(x) = -\frac{(E-\epsilon_2)t_3}{t_2t_1} U_{m-1}(x).
\end{equation}
The eigenspectrum for $m$ up to 20 unit cells ($N$=60) is presented in Fig. \ref{fig:UC3eigenvalues}(b), for the same system as in Fig. \ref{fig:UC3eigenvalues}(a). The bands that are shaped as $N$ increases have widths $\approx 0.51, 0.88, 0.37$ eV, in agreement with the cyclic boundaries, but there are also three eigenvalues that exceed these bands, two of which lie well within the second gap. Generally, our calculations for $u=3$ show that these mid-gap eigenvalues are three at most and their exact number and position depends on complex relations between the TB parameters. Such relations for 3-Toeplitz matrices of order $N=3m+2$ can be found in Ref. \cite{Alvarez:2005}, where the authors obtain in that case two such eigenvalues at most. In our case, where $N = 3m$, we find that, as $N$ increases, these eigenvalues converge at most to three of the zeros of $M_3^{12}\times M_3^{21}$. Specifically, for the parameters of Fig. \ref{fig:UC3eigenvalues}, the positions of the mid-gap eigenvalues, which are shown in Fig. \ref{fig:UC3eigenvalues}(b), converge quickly, as $N$ increases, to $\frac{\epsilon_2+\epsilon_3}{2}\pm \sqrt{(\frac{\epsilon_2-\epsilon_3}{2})^2 + t_2^2}$, i.e. the zeros of $M_3^{12}$, and $\frac{\epsilon_1+\epsilon_2}{2}+ \sqrt{(\frac{\epsilon_1-\epsilon_2}{2})^2 + t_1^2}$, i.e. the maximum of the zeros of $M_3^{21}$.

\subsection{\label{subsec:Evalues4} Unit cell of four sites ( $u=4$, $m={\frac{N}{4}}$)}
\begin{figure} [h!]
\centering
\includegraphics[width=0.9\columnwidth]{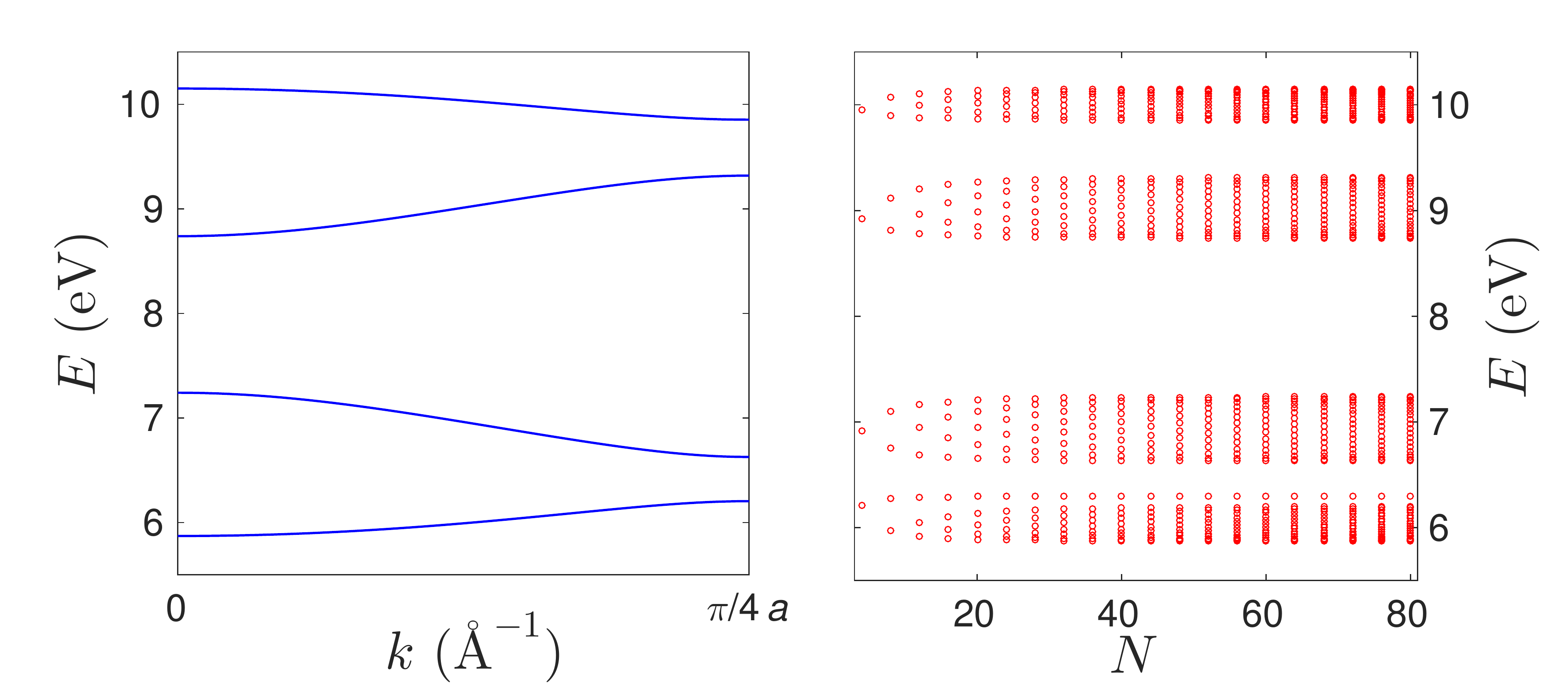}
\caption{(a) Dispersion relation for \textit{cyclic boundaries} and (b) eigenspectrum for \textit{fixed boundaries} for $m$ up to 20 unit cells ($N = 80$) of a 1D periodic lattice  with a unit cell of four sites, with on-site energies $(\epsilon_1,\epsilon_2,\epsilon_3,\epsilon_4) = (7.0,9.0,7.5,8.5)$ eV and hopping integrals $(t_1,t_2,t_3,t_4) = (1.2,0.9,1.0,0.8)$ eV, as determined by Eqs. \eqref{Eq:evsUCTM4pb} and \eqref{Eq:evsUCTM4fb}, respectively.} 
\label{fig:UC4eigenvalues}
\end{figure}

For \textit{cyclic boundaries}, Eq. \eqref{Eq:dispertionrelation} takes the form
\begin{align} \label{Eq:evsUCTM4pb}
\cos(k4a) =& \frac{1}{2} \prod_{i=1}^{4}\frac{(E-\epsilon_i)}{t_i} -\sum_{ijkl}^{\circlearrowright^4}\frac{(E-E_i)(E-E_j)t_k}{2t_it_jt_l}\nonumber \\
&+\frac{t_1t_3}{2t_2t_4} + \frac{t_2t_4}{2t_1t_3},
\end{align}
where $\circlearrowright^4$ denotes all cyclic permutations of the indices $i \neq j \neq k \neq l \in \{1,2,3,4\}$. This dispersion relation is presented in Fig \ref{fig:UC4eigenvalues}(a), for a system with on-site energies $(\epsilon_1,\epsilon_2,\epsilon_3,\epsilon_4) = (7.0,9.0,7.5,8.5)$ eV and hopping integrals $(t_1,t_2,t_3,t_4) = (1.2,0.9,1.0,0.8)$ eV. The four bands that occur for this system have widths $\approx 0.33, 0.61,0.58, 0.30$ eV (from lower to higher energy) and the corresponding gaps are $\approx 0.42, 1.50, 0.53$ eV.

For \textit{fixed boundaries}, it follows from Eq. \eqref{Eq:Eigenvaluesfb2} that the eigenvalues are solutions of
\begin{equation} \label{Eq:evsUCTM4fb}
U_m(x) = \left(-\frac{(E-\epsilon_3)(E-\epsilon_2)t_4}{t_3t_2t_1}+\frac{t_2t_4}{t_1t_3}\right) U_{m-1}(x).
\end{equation}
The eigenspectrum for $m$ up to 20 unit cells ($N$=80) is presented in Fig. \ref{fig:UC4eigenvalues}(b), for the same system as in Fig. \ref{fig:UC4eigenvalues}(a). The bands that are shaped as $N$ increases have widths $\approx 0.33, 0.61, 0.57, 0.30$ eV, in agreement with the cyclic boundaries, but there is also one eigenvalue that exceeds the first band. Generally, there are four mid-gap eigenvalues at most, the exact number and position of which depends on complex relations between the TB parameters. Again, we find that, as $N$ increases, these eigenvalues converge at most to four of the zeros of $M_4^{12}\times M_4^{21}$. Specifically, for the parameters of Fig. \ref{fig:UC4eigenvalues}, the position of the mid-gap eigenvalue, which is shown in Fig. \ref{fig:UC4eigenvalues}(b), converges quickly, as $N$ increases, to the minimum zero of $M_4^{21}$.


\section{\label{sec:DOS} Density of states}
The DOS is defined as the number of states that are available for occupation at an infinitesimal energy interval, $g(E)=\dv{\nu}{E}$. For a periodic system with cyclic boundaries, it can be obtained with the help of the UCTM as
\begin{equation} \label{Eq:DOSdef}
g(E) = \frac{m}{\pi}\dv{E}\abs{\acos(z)} = \frac{m}{\pi}\frac{\abs{\dv{z}{E}}}{\sqrt{1-z^2}},
\end{equation}
where no spin degeneracies are taken into account. It is obvious that $2u$ VHS are expected at the points where $z(E) =\pm1$, i.e. at the energies that lie at the center and edges of the  $1^\text{st}$ BZ. Below, we provide analytical expressions of the DOS for systems with unit cell of 1, 2, 3, and 4 sites, as the simplest examples of the generic periodic system. For large $N$ the DOS for cyclic boundaries converges with the DOS for fixed boundaries.

\subsection{\label{subsec:DOS1} Unit cell of one site ($u=1$, $m=N$)}
\begin{figure} [h!]
\centering
\includegraphics[width=0.9\columnwidth]{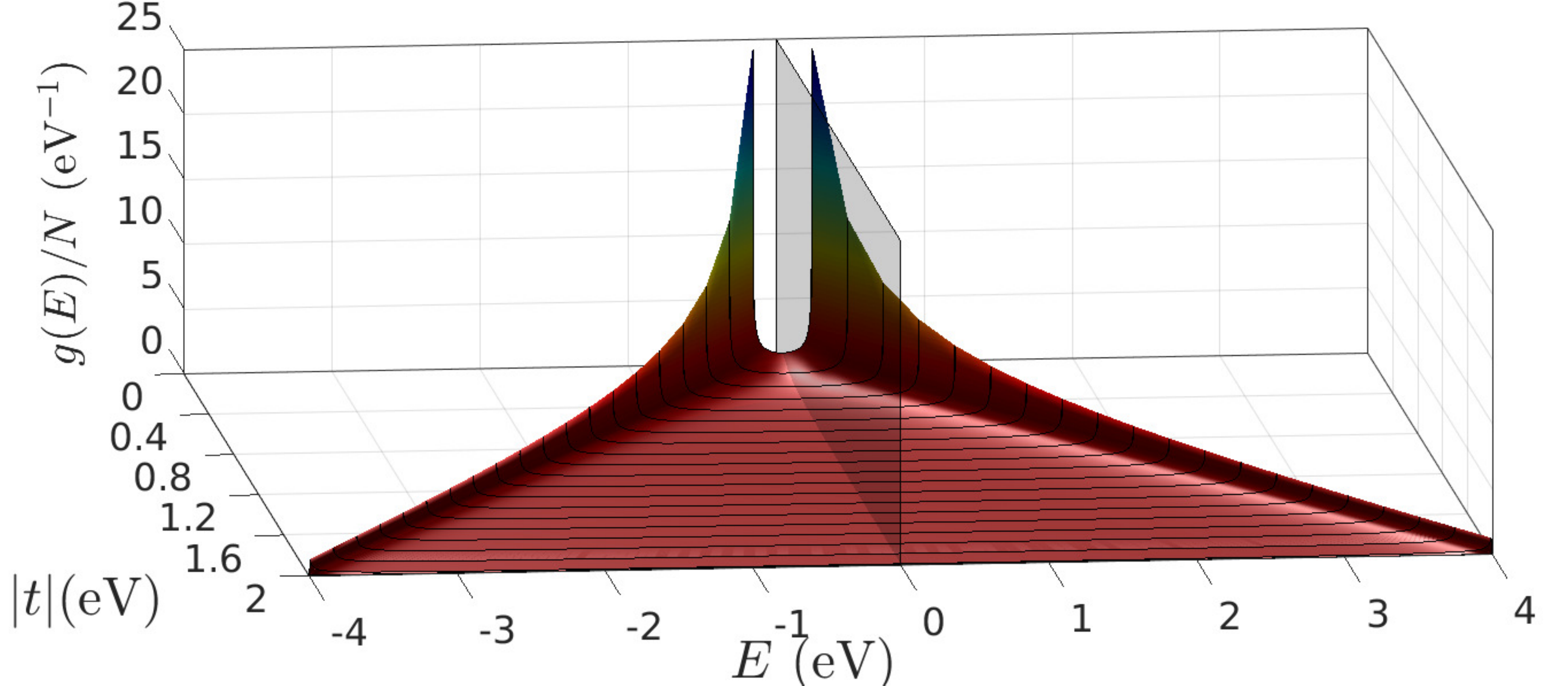}
\caption{DOS (over $N$) of the 1D periodic lattice with a unit cell of one site (Eq. \eqref{Eq:DOS1}), for a fixed on-site energy $\epsilon_1 = 0$, as a function of the magnitude of the hopping integral $\abs{t_1}$.} 
\label{fig:UC1dos}
\end{figure}

Eq. \eqref{Eq:DOSdef} takes the form~\cite{LCMKLTTS:2016}
\begin{equation} \label{Eq:DOS1}
g(E) = \frac{N}{\pi\sqrt{4t^2_1-(E-\epsilon_1)^2}}.
\end{equation}
Two VHS occur at $E = \epsilon_1 \pm 2\abs{t_1}$. The DOS (over the total number of sites $N$) for $u=1$ is presented in Fig. \ref{fig:UC1dos}, for a fixed on-site energy, $\epsilon_1 = 0$, where the effect of alternating the magnitude of the hopping integral, $\abs{t_1}$, is also demonstrated. Increasing the value of $\abs{t_1}$ linearly increases the bandwidth and reduces the sharpness of the VHS that occur at the band edges.

\subsection{\label{subsec:DOS2} Unit cell of two sites ($u=2$, $m=\frac{N}{2}$)}
\begin{figure} [h!]
\centering
\includegraphics[width=0.9\columnwidth]{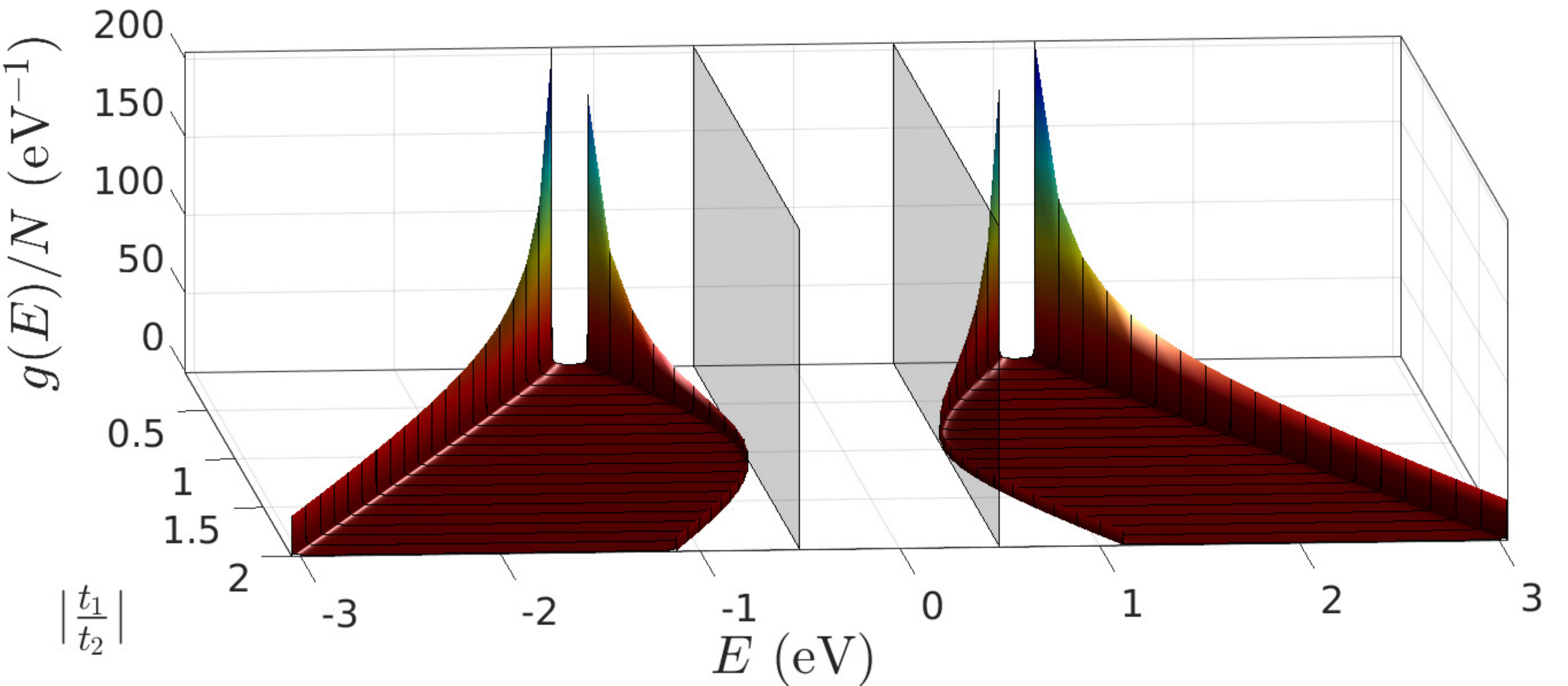}
\caption{DOS (over $N$) of the 1D periodic lattice with a unit cell of two sites (Eq. \eqref{Eq:DOS2}), for fixed on-site energies 
$(\epsilon_1,\epsilon_2) = (-0.5,0.5)$ eV, as a function of the absolute ratio between the hopping integrals, $\abs{\frac{t_1}{t_2}}$.} 
\label{fig:UC2dos}
\end{figure}

Eq. \eqref{Eq:DOSdef} takes the form
\begin{equation} \label{Eq:DOS2}
g(E) = \frac{N}{2\pi}\frac{\abs{2E-\epsilon_1-\epsilon_2}}{\sqrt{4t_1^2t_2^2-\left[(E-\epsilon_1)(E-\epsilon_2)-t_1^2-t_2^2\right]^2}}.
\end{equation}
Four VHS occur at 
$E = \frac{\epsilon_1+\epsilon_2}{2} \pm \sqrt{\left(\frac{\epsilon_1-\epsilon_2}{2}\right)^2 + (t_1 \pm t_2)^2}$. The DOS (over the total number of sites $N$) for $u=2$ is presented in Fig. \ref{fig:UC2dos}, for fixed on-site energies $(\epsilon_1,\epsilon_2) = (-0.5,0.5)$ eV, where the effect of altering the magnitude of the ratio between the hopping integrals, $\abs{\frac{t_1}{t_2}}$, is also demonstrated. The DOS is symmetric around the mean of the on-site energies. Furthermore, the effect of altering $\abs{\frac{t_1}{t_2}}$ on the bandwidths and bandgap, discussed in Subsec. \ref{subsec:Evalues2} (cf. inset of Fig. \ref{fig:typecdisprel3d}) is clearly illustrated. The VHS that are closer to the gap are less sharp. The sharpness of all VHS decreases dramatically as $\abs{\frac{t_1}{t_2}}$ goes from $0$ to 1, and slower for $\abs{\frac{t_1}{t_2}}>1$.

\subsection{\label{subsec:DOS3} Unit cell of three sites ($u=3$, $m=\frac{N}{3}$)}
\begin{figure} [h!]
\centering
\includegraphics[width=0.9\columnwidth]{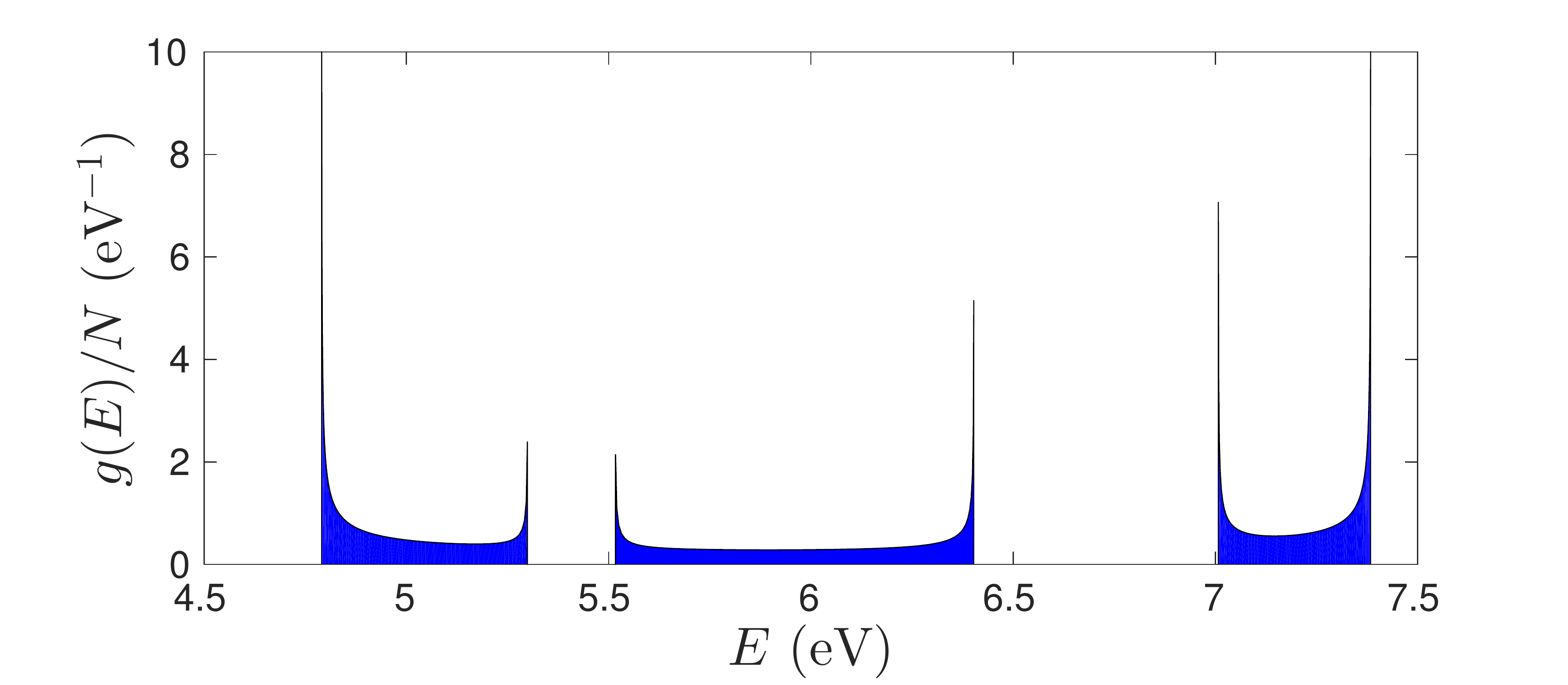}
\caption{DOS (over $N$) of a 1D periodic lattice  with a unit cell of three sites (Eq. \eqref{Eq:DOS3}), with on-site energies $(\epsilon_1,\epsilon_2,\epsilon_3) = (6.3,5.8,6.1)$ eV and hopping integrals $(t_1,t_2,t_3) = (0.5,0.6,0.8)$ eV.} 
\label{fig:UC3dos}
\end{figure}

Eq. \eqref{Eq:DOSdef} takes the form
\begin{equation} \label{Eq:DOS3}
g(E) = \frac{N}{3\pi}\frac{\abs{\displaystyle\sum_{ijk}^{\circlearrowright^3}\frac{(E-\epsilon_i)(E-\epsilon_j)}{2t_it_jt_k}-\frac{t_j}{2t_kt_i}}}{\sqrt{1-\left[\frac{1}{2} \displaystyle\prod_{i=1}^{3}\frac{(E-\epsilon_i)}{t_i} -\sum_{ijk}^{\circlearrowright^3}\frac{(E-\epsilon_i)t_j}{2t_kt_i}\right]^2}}.
\end{equation}
In Fig. \ref{fig:UC3dos}, we illustrate the DOS of a system with three sites per unit cell, for the parameters used in Fig. \ref{fig:UC3eigenvalues}, as obtained by Eq. \eqref{Eq:DOS3}. As a careful comparison of Figs. \ref{fig:UC3eigenvalues}(b) and  \ref{fig:UC3dos} suggests, the relative sharpness of the VHS as well as the position of each band's DOS minimum can be expected from the density of points in the eigenspectrum, as $N$ increases. This is another demonstration that in the large $N$ limit, the boundary effects play insignificant role in the electronic structure of the system, since the mid-gap eigenvalues that occur for fixed boundaries are negligible in number compared to those that lie within the bands.

\subsection{\label{subsec:DOS4} Unit cell of four sites ($u=4$, $m=\frac{N}{4}$)}
\begin{figure} [h!]
\centering
\includegraphics[width=0.9\columnwidth]{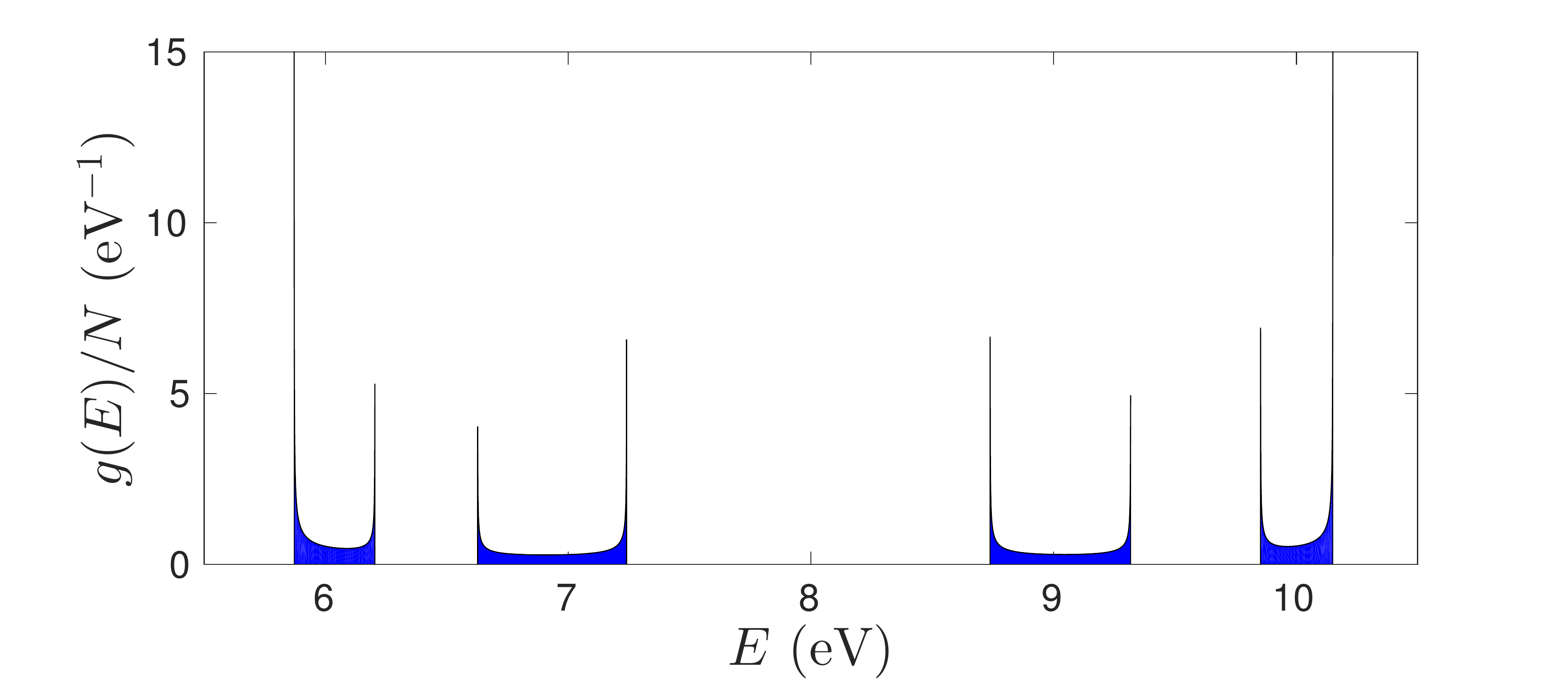}
\caption{DOS (over $N$) of a 1D periodic lattice  with a unit cell of four sites (Eq. \eqref{Eq:DOS4}), with on-site energies $(\epsilon_1,\epsilon_2,\epsilon_3,\epsilon_4) = (7.0,9.0,7.5,8.5)$ eV and hopping integrals $(t_1,t_2,t_3,t_4) = (1.2,0.9,1.0,0.8)$ eV.} 
\label{fig:UC4dos}
\end{figure}

Eq. \eqref{Eq:DOSdef} takes the form
\begin{widetext}
\begin{equation} \label{Eq:DOS4}
g(E) = \frac{N}{4\pi}\frac{\displaystyle\abs{\sum_{ijkl}^{\circlearrowright^4}\frac{(E-\epsilon]_i)(E-\epsilon_j)(E-\epsilon_k)}{2t_it_jt_kt_l}-\frac{(2E-\epsilon_i-\epsilon_j)t_k}{2t_it_jt_l}}}{\sqrt{1-\left[\displaystyle\frac{1}{2} \prod_{i=1}^{4}\frac{(E-\epsilon_i)}{t_i} -\sum_{ijkl}^{\circlearrowright^4}\frac{(E-\epsilon_i)(E-\epsilon_j)t_k}{2t_it_jt_l}+ \frac{t_1t_3}{2t_2t_4} + \frac{t_2t_4}{2t_1t_3}\right]^2}}.
\end{equation}
\end{widetext}
In Fig. \ref{fig:UC4dos}, we illustrate the DOS of a system with four sites per unit cell, for the parameters used in Fig. \ref{fig:UC4eigenvalues}, as obtained by Eq. \eqref{Eq:DOS4}.

\section{\label{sec:TC} Transmission coefficient}
The transmission coefficient describes the probability that a carrier, incident to a quantum system, transmits through its eigenstates. To obtain the TC of a 1D TB lattice, we connect the ends of the system under examination to semi-infinite homogeneous leads, which play the role of a carrier bath. Hence, sites $[-\infty,0] \cup [N+1,\infty]$ belong to the left and right lead, respectively. Within sites $[1,N]$ lies the periodic lattice under examination, which is considered as a perturbation in the homogeneous infinite lead. The coupling between the periodic system and the left and right lead is described by the effective parameters $t_{cL}$ and  $t_{cR}$, respectively. In this article we generally choose them to be different, as a reflection of the difference in coupling of the same material (lead) with different moieties (the end sites of the system). The effect of this asymmetry in coupling will be discussed in detail below. The leads' band lies in the energy interval 
$[\epsilon_m-2\abs{t_m},\epsilon_m+2\abs{t_m}]$, where $\epsilon_m$ is the on-site energy of the leads and $t_m$ the hopping integral between the leads' sites, and their dispersion relation at zero bias is similar to Eq. \eqref{Eq:evsUCTM1pb}. Hence, the energy center and bandwidth of the leads are $\epsilon_m$ and $4\abs{t_m}$, respectively. If we imagine the lead as a homogeneous system with one electron per site, then the band is half-filled (hence the leads are metallic), and $\epsilon_m$ is the Fermi energy of the metal.
The GTM of the whole lead-system-lead complex is $\tilde{M}_R P_R \tilde{M}_N P_L \tilde{M}_L$, where $\tilde{M}_{L/R}$ represents the sites of the ideal, cyclically bounded left/right lead, and
\begin{equation} \label{Eq:PLR}
P_R\begin{pmatrix}
\frac{t_u}{t_{cR}} & 0 \\ 0 &  \frac{t_{cR}}{t_m}
\end{pmatrix}, \quad
P_L =
\begin{pmatrix}
\frac{t_m}{t_{cL}} & 0 \\ 0 & \frac{t_{cL}}{t_u}
\end{pmatrix}
\end{equation}
are the matrices that describe the coupling of the three subsystems. 
The waves at the left and the right lead can be expanded as
\begin{equation} \label{Eq:waves}
\psi_{\{n\}\leq 1} = e^{iq_Lna} + r e^{-iq_Lna}, \quad \psi_{\{n\}\geq N} = t e^{iq_Rna},
\end{equation}
where $q_{L/R}$ is the wavenumber (at zero bias, $q_L = q_R = q$). 
We have assumed, without any loss of generality, that the incident waves come from the left and we have normalized their amplitude. 
Hence, the TC is defined as $T(E) = \abs{t}^2$. For a periodic system, the GTM of the perturbation's region obeys to the equation
\begin{equation} \label{Eq:MNs}
\begin{pmatrix}
\psi_{N+1} \\ \psi_N
\end{pmatrix} = P_R M_u^m P_L
\begin{pmatrix}
\psi_{1} \\ \psi_0
\end{pmatrix}.
\end{equation}
After manipulations, we find that the TC at zero bias is
\begin{equation} \label{Eq:TCdef}
T(E) = \frac{1}{1+\Lambda(E)},
\end{equation}
\begin{equation} \label{Eq:Lambdadef}
\Lambda(E) = \frac{\left[W_N(E) + X_N^+(E) \cos(q a)\right]^2}{4 \sin[2](q a)} + \frac{X_N^-(E)^2}{4}.
\end{equation}
For a periodic system,
\begin{equation} \label{Eq:Wdefper}
W_N(E) = U_m(z) \omega - U_{m-1}(z) M_u^{22} (\omega + \omega^{-1}) + U_{m-2}(z) \omega^{-1},
\end{equation}
\begin{equation} \label{Eq:Xdefper}
X_N^\pm(E) = U_{m-1}(z) X_u^\pm(E),
\end{equation}
\begin{equation} \label{Eq:Xu}
X_u^\pm(E) = (M_u^{12} \chi \pm M_u^{21} \chi^{-1}). 
\end{equation}
\begin{equation} \label{Eq:omegadef}
\omega = \frac{t_m \; t_u}{t_{cR} \; t_{cL}},
\end{equation}
which is included only in $W_N(E)$, expresses the coupling \textbf{strength}, in means of the deviation of the real coupling of the system to the leads from the ideal coupling (in which the system and the leads are interconnected as if they were connected to themselves). The term
\begin{equation} \label{Eq:chidef}
\chi = \frac{t_{cL}}{t_{cR}},
\end{equation}
which is included only in $X_N^\pm(E)$, expresses the coupling \textbf{asymmetry}, i.e. the difference in coupling strength between the leads and the left/right end of the system. The maxima/minima of TC are the minima/maxima of $\Lambda(E)$, which is a polynomial of energy of degree $N$. In the following Subsections, we discuss in detail the effect of the coupling strength and asymmetry, as well as the lead properties, to the transmission profile of any periodic system. In all cases considered below, we will assume that the bandwidth of the leads, as determined by $\abs{t_m}$ is such that it contains all the eigenstates of the system, in order to gain the full picture of the transmission profiles.

\subsection{\label{subsec:idealcoupling}Ideal coupling}
\begin{figure} [h!]
\centering
\includegraphics[width=\columnwidth]{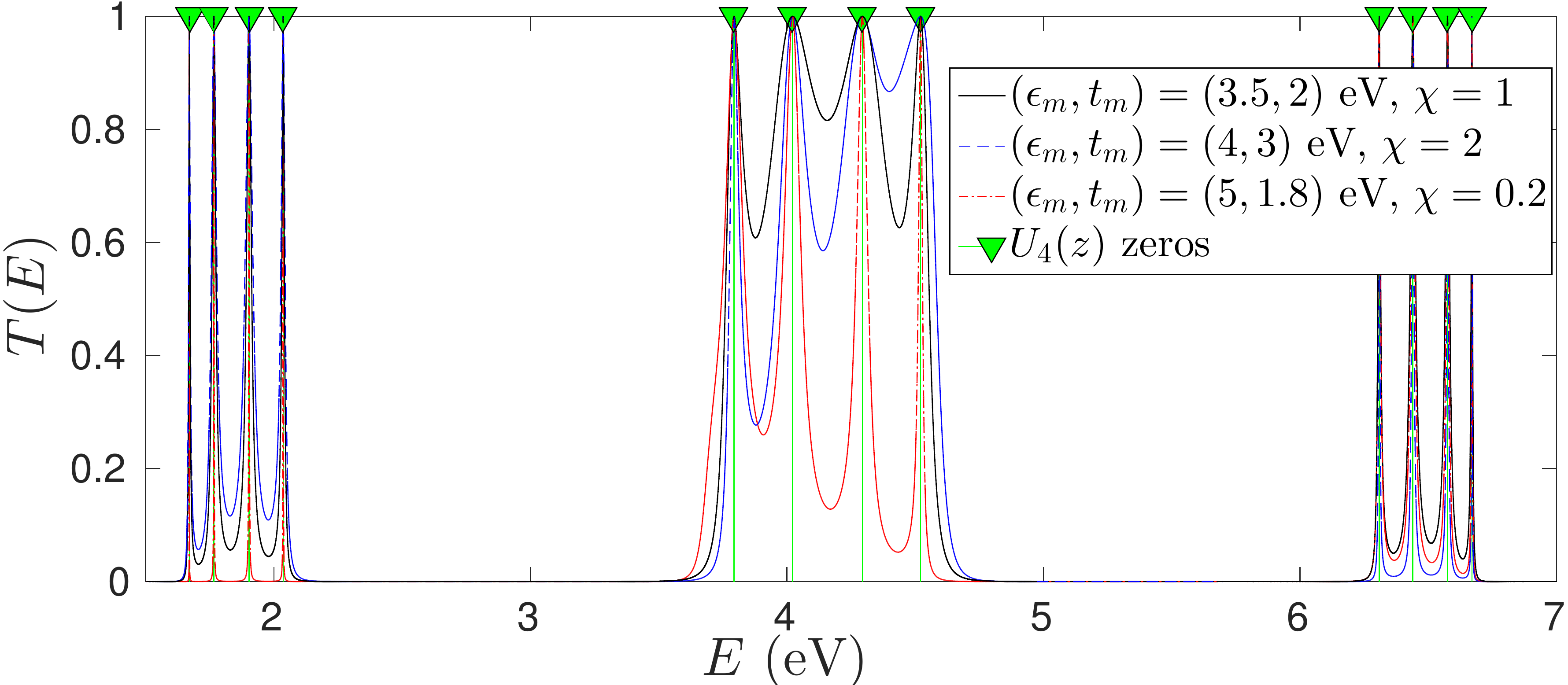}
\caption{Transmission coefficient of a periodic system with $u = 3$ and $m = 5$ ($N = 15$), for the ideal coupling condition $\omega = 1$. The TB parameters of the system are $(\epsilon_1,\epsilon_2, \epsilon_3) = (3, 4, 5.5)$ eV and $(t_1,t_2,t_3) = (1, 0.8, 1.5)$ eV. The energies that correspond to full transmission do not depend on the energy center and bandwidth of the leads or the coupling asymmetry; they depend solely on the energy structure of the system, their number is $(m-1)u= 12$, and they are given by the zeros of $U_4(z)$.}
\label{fig:TCWMomega1}
\end{figure}

For ideal coupling between the leads and the system, i.e. when $\abs{\omega} = \abs{\omega^{-1}} = 1$, the recurrent formula Eq. \eqref{Eq:ChebyshevUrec} results in $W_N = \text{sgn}(\omega) U_{m-1}(z) (M_u^{11}-M_u^{22})$. Hence, 
\begin{align} \label{Eq:Lambdaideal}
\Lambda(E) = \biggl\{ &\frac{\left[\text{sgn}(\omega) (M_u^{11}-Mu^{22}) + X_u^+(E) \cos(\kappa a)\right]^2}{4 \sin[2](\kappa a)} \nonumber\\
&+ \frac{X_u^-(E)^2}{4}\biggr\} U_{m-1}(z)^2.  
\end{align}
From Eq. \eqref{Eq:Lambdaideal} it is obvious that if $U_{m-1}(z) = 0$, then $T(E) = 1$. Hence, for ideal coupling, there are $(m-1)u = N-u$ energies in which transmission becomes full. These energies lie in the zeros of $U_{m-1}(z)$, i.e. they are the solutions of $z(E) = \cos(\frac{\mu \pi}{m})$, $\mu = 1,2,\dots,m-1$. Hence their position depends solely on the energy structure of the system and not on the energy center/bandwidth of the leads or the coupling asymmetry. This is depicted in Fig. \ref{fig:TCWMomega1}, for a periodic system with $u = 3$ sites per unit cell and $m = 5$ unit cells ($N = 15$). Finally, we should mention that this full transmission condition that occurs for ideal coupling is solely a result of the periodicity of the system, i.e. the fact that the elements of $\tilde{M}_N$ are related to the Chebyshev polynomials of the second kind. The effect of the energy center and bandwidth of the leads will be separately addressed below.

\subsubsection{\label{subsubsec:idealcouplingchi} The role of coupling asymmetry}
\begin{figure} [h!]
\centering
\includegraphics[width=\columnwidth]{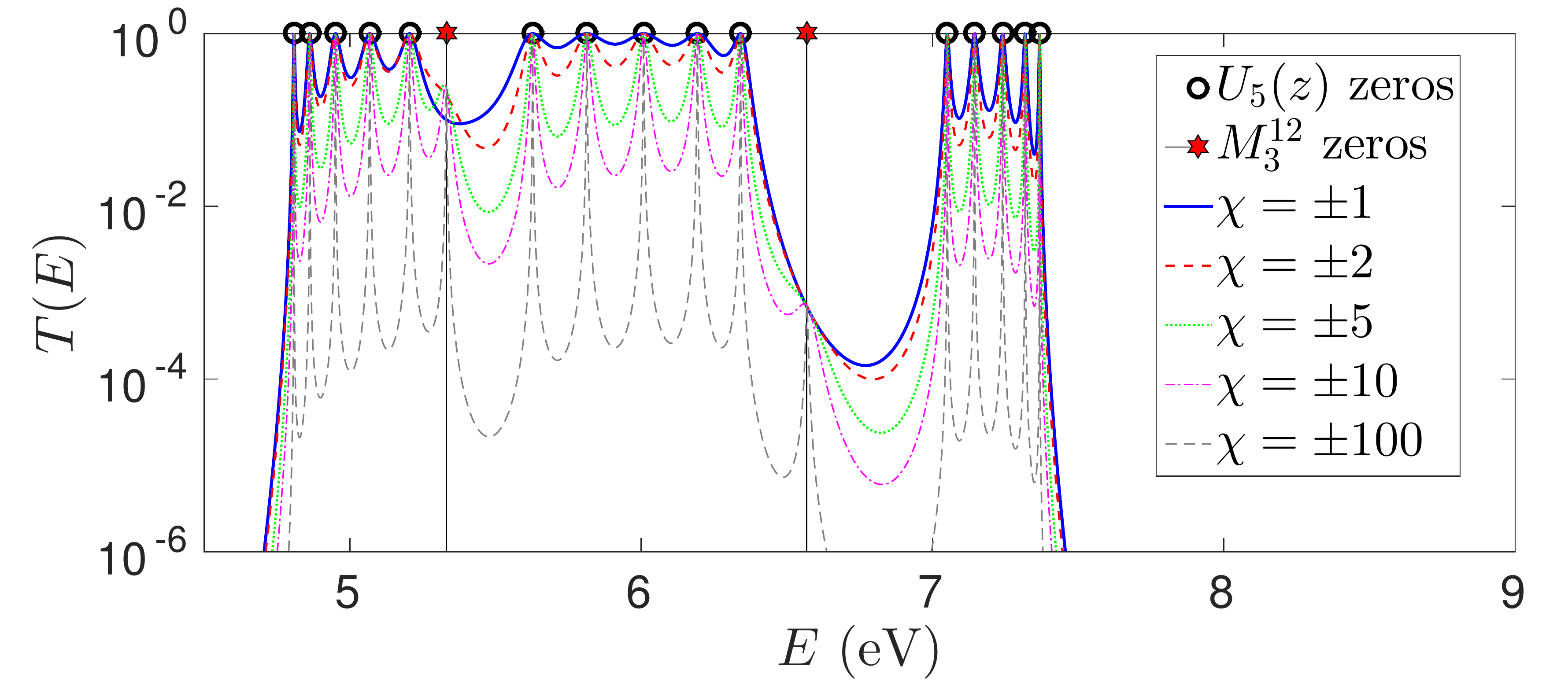}\\
\includegraphics[width=\columnwidth]{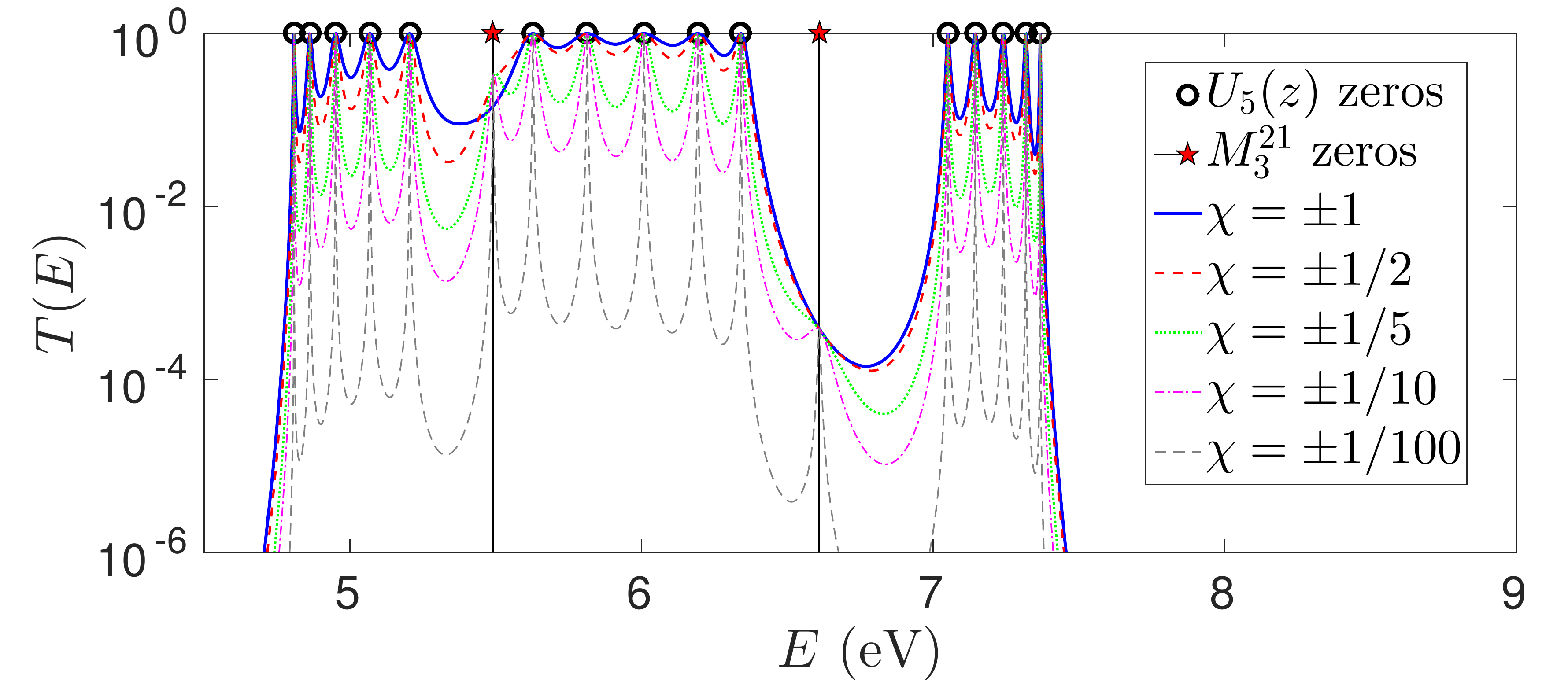}
\caption{Dependence of the transmission coefficient on the coupling asymmetry $\chi$ for a periodic system with $u = 3$ sites per unit cell and $m = 6$ unit cells ($N = 18$), for the ideal coupling condition $\abs{\omega} = 1$. The TB parameters of the system are  $(\epsilon_1,\epsilon_2, \epsilon_3) = (6.3, 5.8, 6.1)$ eV and $(t_1,t_2,t_3) = (0.5, 0.6, 0.8)$ eV. The on-site energy and hopping integral of the leads are chosen $E_m = 6$ eV, $t_m = 2$ eV. [Top] $T(E)$ for $\abs{\chi} > 1$. [Bottom] $T(E)$ for $\abs{\chi} < 1$. The transmission becomes full in $(m-1)u= 15$ energies, which are given by the zeros of $U_{5}(z)$. As $\abs{\chi}$ increases (decreases) above (below) $1$, $u-1$ secondary peaks occur, the energy of which is related to the zeros of $M_u^{12}$  ($M_u^{21}$). }
\label{fig:TCWMomega1chiandinvchi}
\end{figure}

The role of coupling asymmetry is depicted 
in Fig.~\ref{fig:TCWMomega1chiandinvchi}, for the system 
of Fig.~\ref{fig:UC3eigenvalues} with $m = 6$ unit cells ($N = 18$). 
We notice that increasing (decreasing) $\abs{\chi}$ above (below) $1$ leads to a significant decrease in the lower envelope of $T(E)$ and a sharpening of the transmission peaks, which reflects the enhancement of backscattering effects. Furthermore, such an increase (decrease) of $\abs{\chi}$ above (below) $1$ leads to the appearance of $u-1$ secondary peaks, the positions of which are related to the zeros of $M_u^{12}$  ($M_u^{21}$). These peaks occur because this increase (decrease) of  $\abs{\chi}$ leads to the domination of the terms $X^\pm(E)$ inside the curly brackets of Eq. \eqref{Eq:Lambdaideal}.

\subsubsection{\label{subsubsec:idealcouplingleads} The role of the leads}
\begin{figure} [h!]
\centering
\includegraphics[width=0.9\columnwidth]{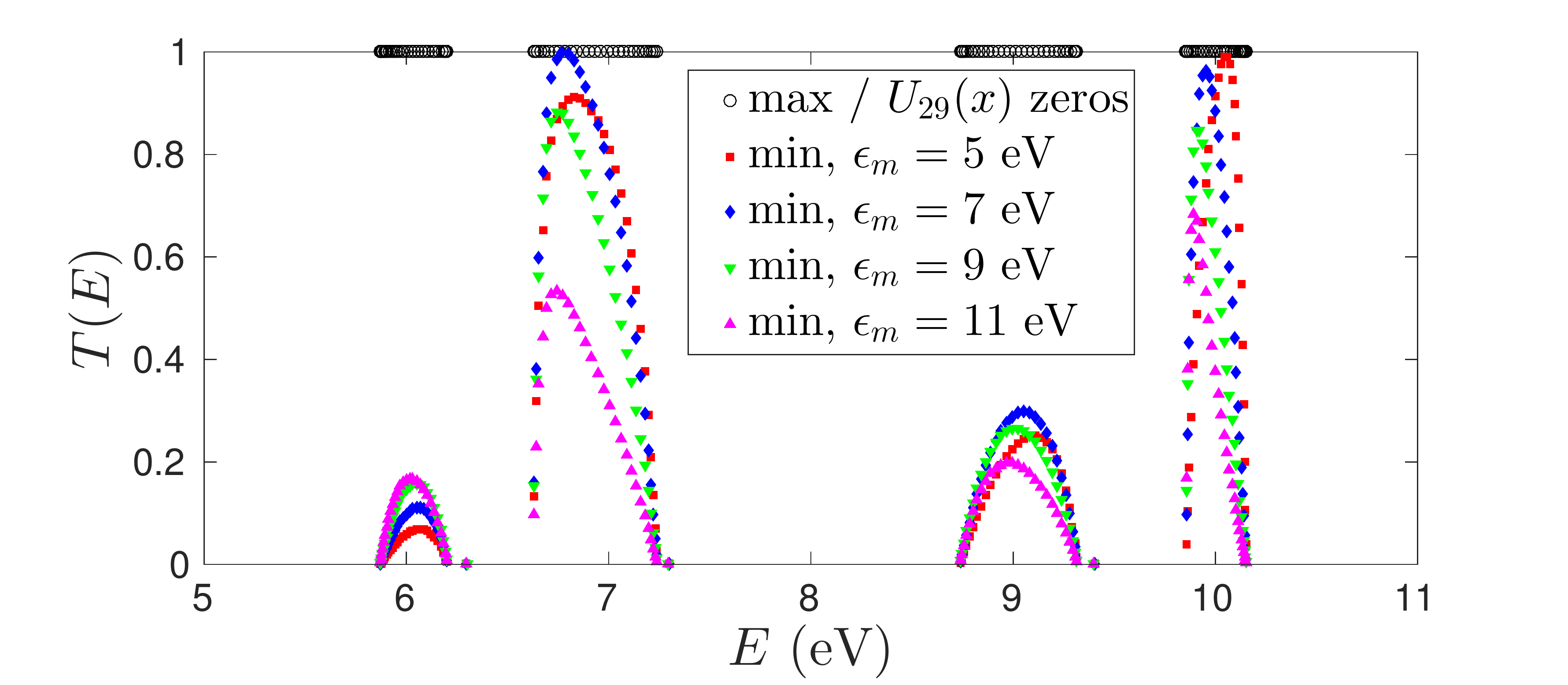}\\
\includegraphics[width=0.9\columnwidth]{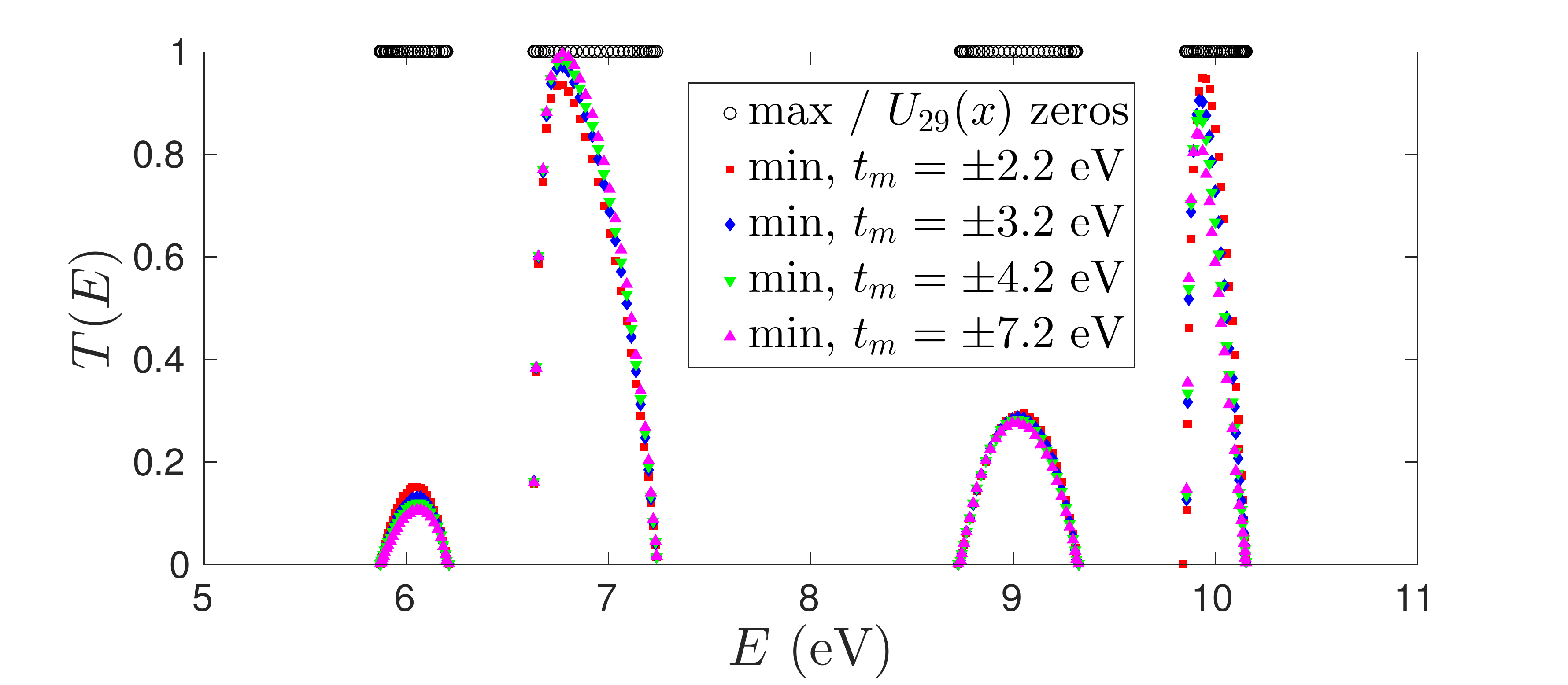}
\caption{Dependence of the transmission coefficient on the leads properties for a periodic system with $u = 4$ sites per unit cell and $m = 30$ unit cells ($N = 120$), for the ideal coupling condition and symmetric coupling $\abs{\omega} = \abs{\chi} = 1$. The TB parameters of the system are 
$(\epsilon_1,\epsilon_2, \epsilon_3,\epsilon_4) = (7, 9, 7.5, 8.5)$ eV and 
$(t_1,t_2,t_3,t_4) = (1.2, 0.9, 1, 0.8)$ eV. 
[Top] $T(E)$ for fixed bandwidth $4\abs{t_m}$ ($t_m = 3$ eV) and varying energy center $\epsilon_m$. 
[Bottom] $T(E)$ for fixed energy center ($\epsilon_m = 8$ eV) and varying $t_m$. The transmission becomes full in $(m-1)u= 116$ energies, which are given by the zeros of $U_{29}(z)$. }
\label{fig:TCWMomega1leads}
\end{figure}

Let us take as example, the system of Fig. \ref{fig:UC4eigenvalues} with $m = 30$ unit cells ($N = 120$) and suppose symmetric coupling, $\abs{\chi} = 1$.  For a fixed bandwidth of the leads, $4\abs{t_m}$, we observe that 
changing the energy center of the leads, $\epsilon_m$, has a significant effect on the lower envelopes of $T(E)$, i.e. the ``curve'' shaped by the local minima of $T(E)$ (Fig.~\ref{fig:TCWMomega1leads} [Top]). Generally, the increase of $\epsilon_m$ leads to a shift of the maxima of the lower envelopes to smaller energies. If we fix $\epsilon_m$, we find that the increase of the leads' bandwidth,  $4\abs{t_m}$, has significantly less effect on the transmission profiles, compared to the increase of $\epsilon_m$ (Fig.~\ref{fig:TCWMomega1leads} [Bottom]).

\subsection{\label{subsec:strongweakcoupling}Strong or weak coupling}
\begin{figure} [h!]
\centering
\includegraphics[width=\columnwidth]{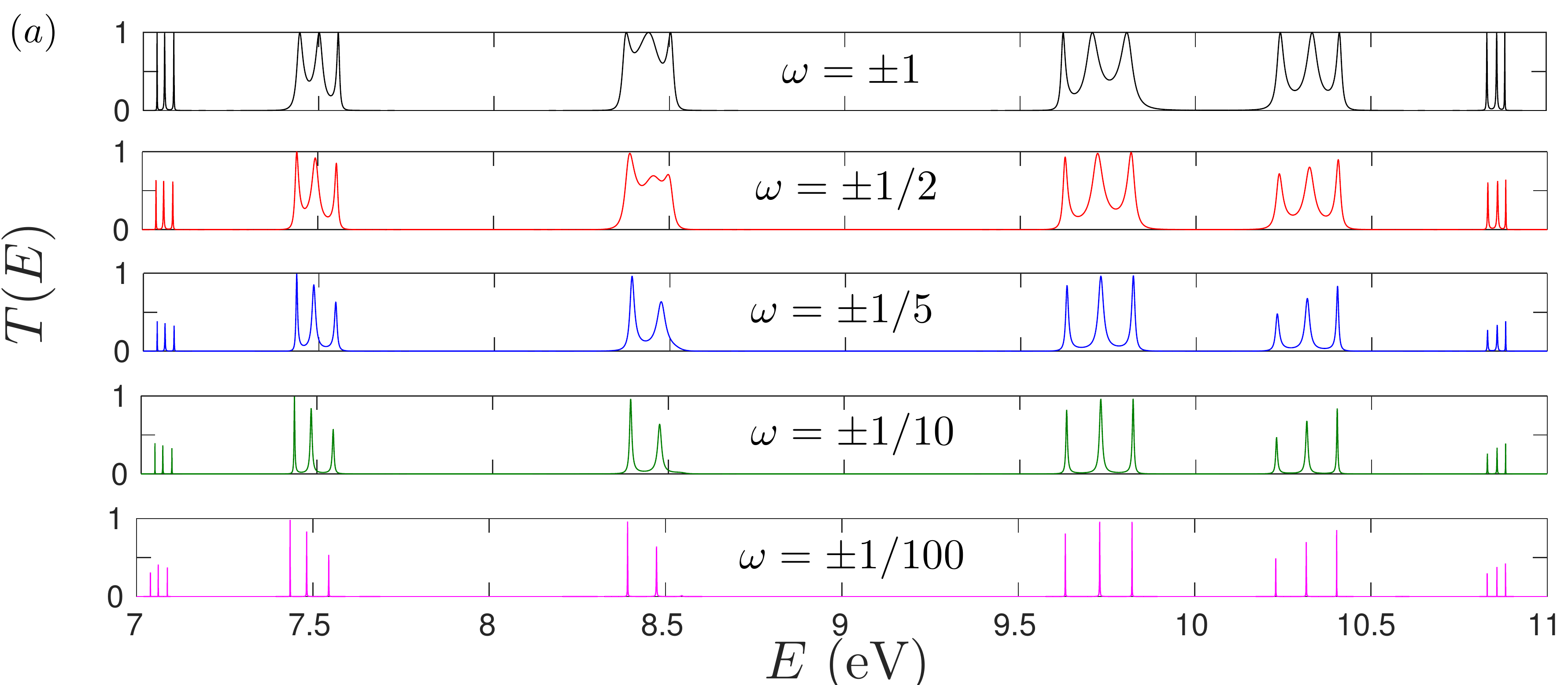}\\
\includegraphics[width=\columnwidth]{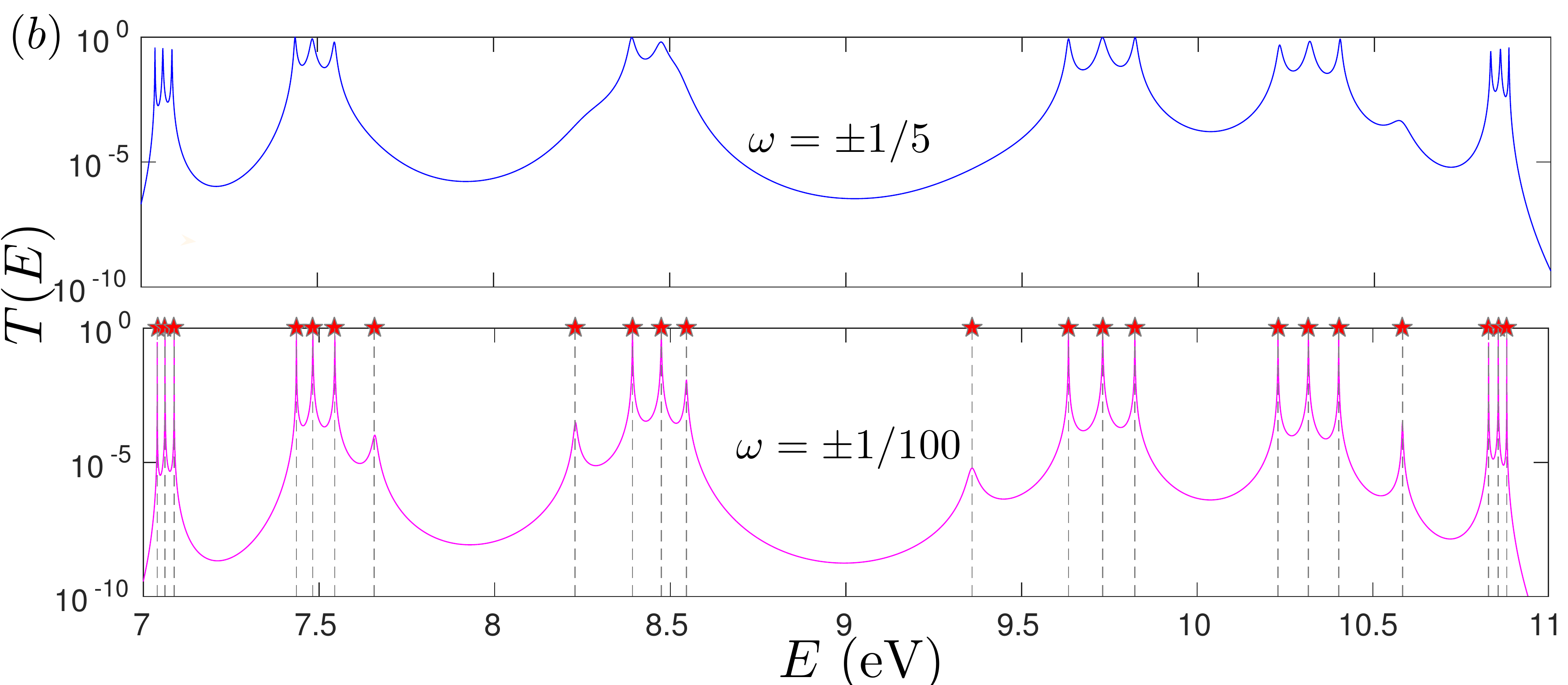}
\caption{(a) Dependence of the transmission coefficient $T(E)$ on the coupling strength factor $\omega$, in the strong coupling regime ($\abs{\omega^{-1}} > 1$), for symmetric coupling ($\abs{\chi} = 1$), for a periodic system with $u = 6$ and $m = 4$ ($N = 24$). (b) Transmission coefficients for strong ($\abs{\omega} = \frac{1}{5}$) and very strong ($\abs{\omega} = \frac{1}{100}$) coupling in logarithmic scale. As the coupling strength $\abs{\omega^{-1}}$ increases above the ideal coupling condition, $N-2$ peaks arise. For very strong coupling, the peaks' position are determined by the zeros of Eq. \eqref{Eq:WNWMinvomegagg}, depicted in dashed lines, in the bottom panel of (b).}
\label{fig:TCWMinvomegalarge}
\end{figure}

\begin{figure} [h!]
\centering
\includegraphics[width=\columnwidth]{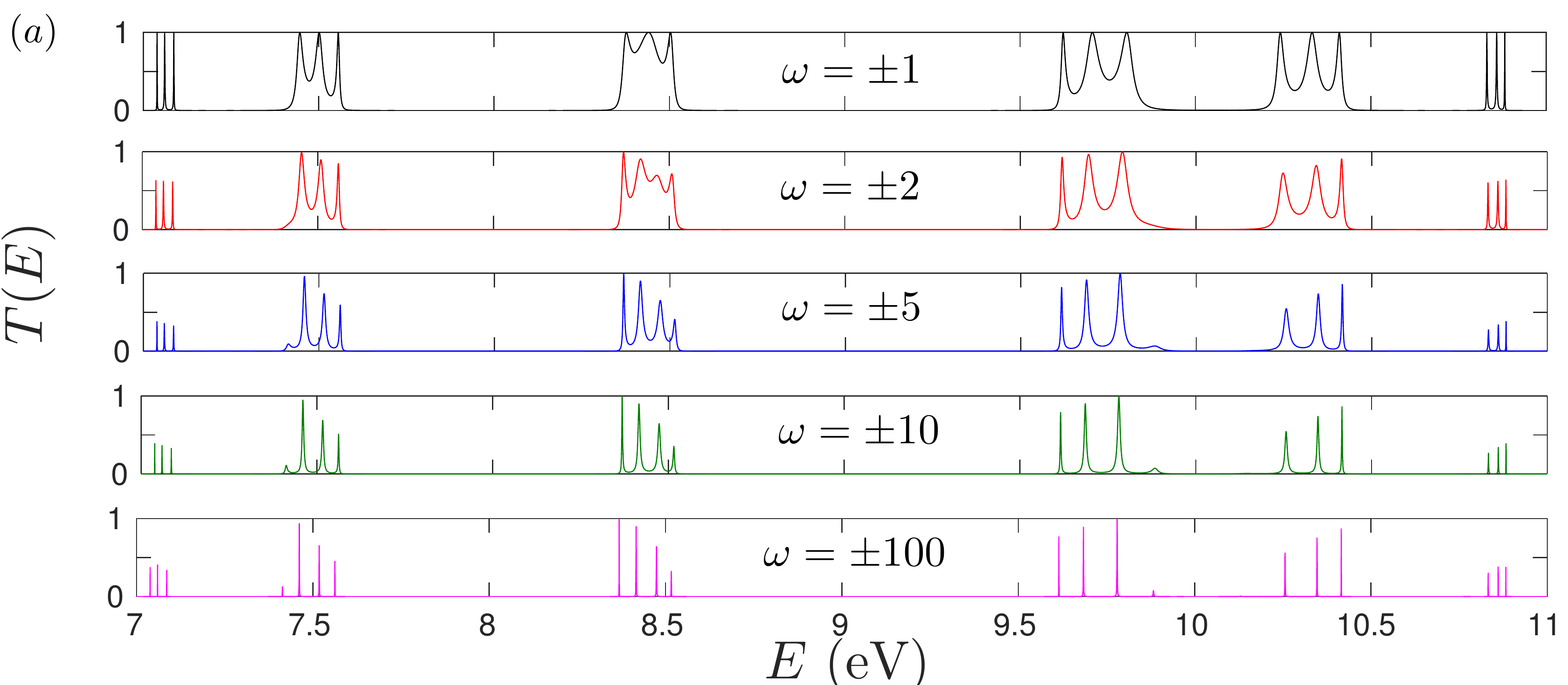}\\
\includegraphics[width=\columnwidth]{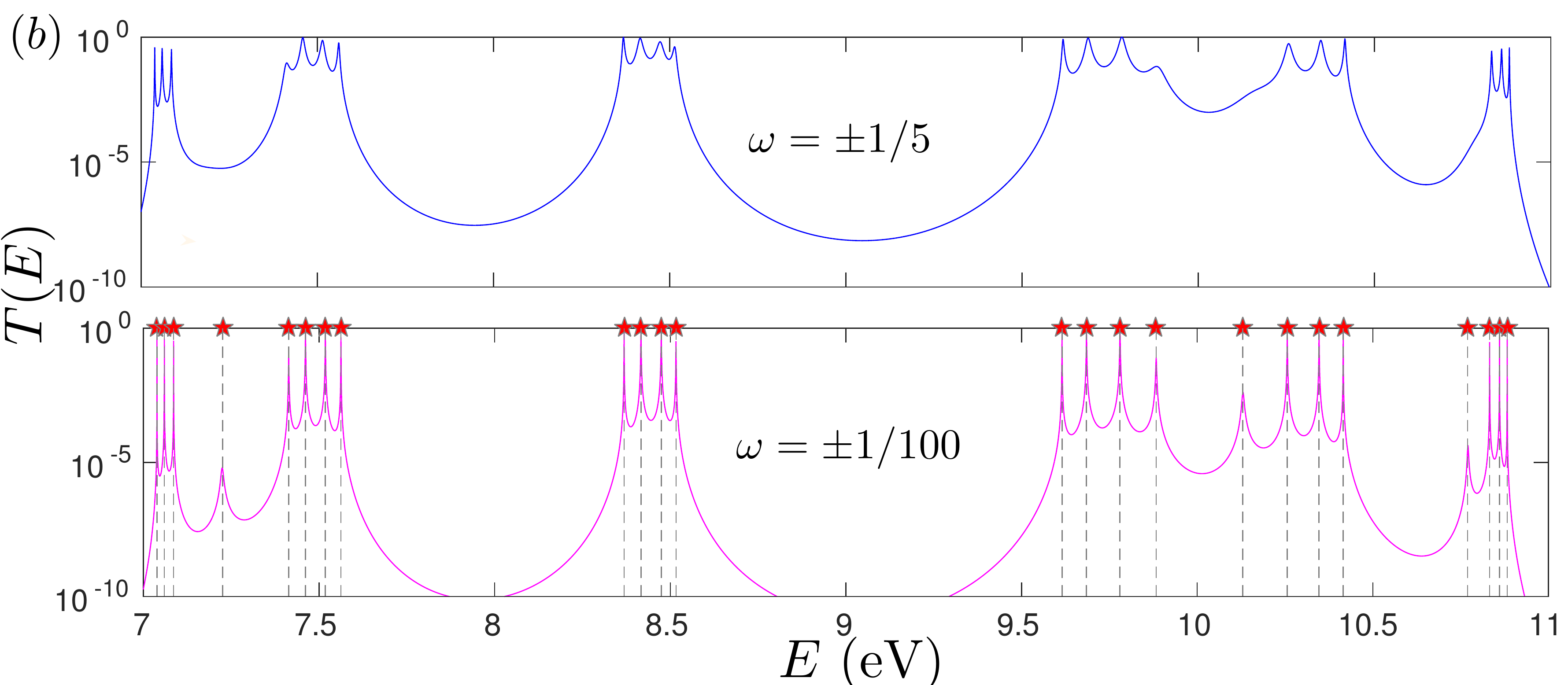}
\caption{(a) Dependence of the transmission coefficient $T(E)$ on the coupling strength factor $\omega$, in the weak coupling regime ($\abs{\omega} > 1$), for symmetric coupling ($\abs{\chi} = 1$), for a periodic system with $u = 6$ and $m = 4$ ($N = 24$). (b) Transmission coefficients for weak ($\abs{\omega} = 5$) and very weak ($\abs{\omega} = 100$) coupling in logarithmic scale. As the coupling weakness $\abs{\omega}$ increases above the ideal coupling condition, $N$ peaks arise. For very weak coupling, the peaks' position are determined by the zeros of Eq. \eqref{Eq:WNWMomegagg}, depicted in dashed lines, in the bottom panel of (b).}
\label{fig:TCWMomegalarge}
\end{figure}

From Eq. \eqref{Eq:Wdefper} it follows that in the very strong coupling regime ($\abs{\omega^{-1}} \gg 1$) 
\begin{equation} \label{Eq:WNWMinvomegagg}
W_N(E) \simeq \omega^{-1} (U_{m-2}(z) - U_{m-1}(z) M_u^{22}),
\end{equation}
while, in the very weak coupling regime ($\abs{\omega} \gg 1$)
\begin{equation} \label{Eq:WNWMomegagg}
W_N(E) \simeq \omega (U_{m}(z) - U_{m-1}(z) M_u^{22}).
\end{equation}
If the coupling is strong or weak and symmetric, the term $W_N(E)$ becomes dominant in Eq. \eqref{Eq:Lambdadef}. Hence, for very strong or weak symmetric coupling, the transmission peaks occur in the region of the zeros of $W_N(E)$, as given by Eq. \eqref{Eq:WNWMinvomegagg} or \eqref{Eq:WNWMomegagg}, hence,  $N-2$ or $N$ peaks are expected. A depiction of the effect of the decrease  (increase) of coupling strength factor $\omega$ below (above) the ideal coupling condition, for symmetric coupling, is presented in Fig. \ref{fig:TCWMinvomegalarge} (\ref{fig:TCWMomegalarge}), for a system with $u = 6$ and $m = 4$ ($N = 24$). The TB parameters of the system are $(\epsilon_1,\epsilon_2, \epsilon_3, \epsilon_4,\epsilon_5, \epsilon_6)$ $= (10, 8.3, 9.7, 8.8, 9.1, 8)$ eV and $(t_1,t_2,t_3,t_4,t_5,t_6) = (0.9, 1, 0.8, 0.6, 1.1, 0.7)$ eV. The leads' parameters are $\epsilon_m = 9$ eV, $t_m = 2$ eV.

\subsubsection{\label{subsubsec:strongweakcouplingchi} The role of coupling asymmetry}
\begin{figure} [h!]
\centering
\includegraphics[width=0.9\columnwidth]{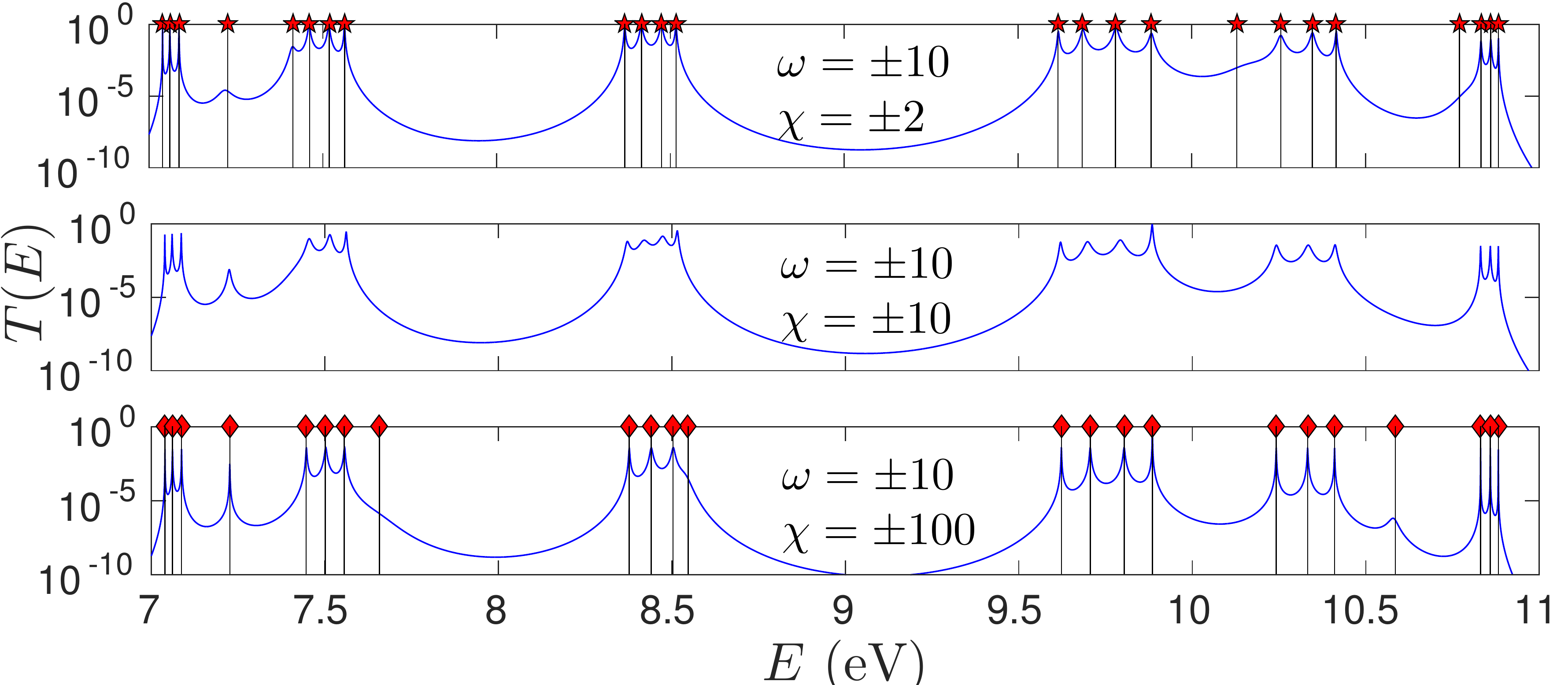}
\caption{Dependence of the transmission coefficient $T(E)$ on the coupling asymmetry $\abs{\chi}$, in the weak coupling regime ($\abs{\omega} = 10$), for $\abs{\chi} = 2$ (top), $10$ (middle), $100$ (bottom), for a periodic system with $u = 6$ and $m = 4$ ($N = 24$). For $\abs{\chi} \ll \abs{\omega}$, the transmission peaks occur in the region of $N$ the zeros of Eq. \eqref{Eq:WNWMomegagg}, which are also depicted in the top panel. For $\abs{\chi} \gg \abs{\omega}$, the transmission peaks occur in the region of the $N-1$ zeros of $U_{m-1}(z)M_u^{12}$, which are also depicted in the bottom panel. The case in which the asymmetry of coupling is comparable to is weakness, is an intermediate regime.}
\label{fig:TCWMomegalargeandasymmetry}
\end{figure}

If the coupling is strong (weak) and asymmetric, three cases can be distinguished: (a) If the coupling asymmetry is significantly smaller than the coupling strength or weakness [$\max(\abs{\chi},\abs{\chi^{-1}}) \ll \abs{\omega^{-1}}$ for strong coupling or $\max(\abs{\chi},\abs{\chi^{-1}}) \ll \abs{\omega}$ for weak coupling], $W_N(E)$ continues to be dominant in Eq. \eqref{Eq:Lambdadef}, hence, the transmission peaks occur again in the region of the zeros of $W_N(E)$ as given by Eq. \eqref{Eq:WNWMinvomegagg} for strong coupling or by Eq. \eqref{Eq:WNWMomegagg} for weak coupling. (b) If the coupling asymmetry is of comparable magnitude with the coupling strength or weakness [$\max(\abs{\chi},\abs{\chi^{-1}}) \approx \abs{\omega^{-1}}$ for strong coupling or $\max(\abs{\chi},\abs{\chi^{-1}}) \approx \abs{\omega}$ for weak coupling], the peaks position cannot be determined without the full solution of Eq. \eqref{Eq:Lambdadef}. (c) If the coupling asymmetry is significantly larger than the coupling strength or weakness [$\max(\abs{\chi},\abs{\chi^{-1}}) \gg \abs{\omega^{-1}}$ for strong coupling or $\max(\abs{\chi},\abs{\chi^{-1}}) \gg \abs{\omega}$ for weak coupling], the terms $X_Ν^\pm(E)$ become dominant in Eq. \eqref{Eq:Lambdadef}, hence the transmission peaks occur in the region of the zeros of  $U_{m-1}(x)M_u^{12}$ for $\abs{\chi} > 1$ or of $U_{m-1}(z) M_u^{21}$ for $\abs{\chi} < 1$. The peaks number in this case is $N-1$. This behavior is depicted in Fig. \ref{fig:TCWMomegalargeandasymmetry}, in the case of weak and asymmetric coupling, for the same system as in Figs. \ref{fig:TCWMinvomegalarge} and \ref{fig:TCWMomegalarge}.

\subsubsection{\label{subsubsec:strongweakcouplingleads} The role of the leads}
\begin{figure} [h!]
\centering
\includegraphics[width=0.9\columnwidth]{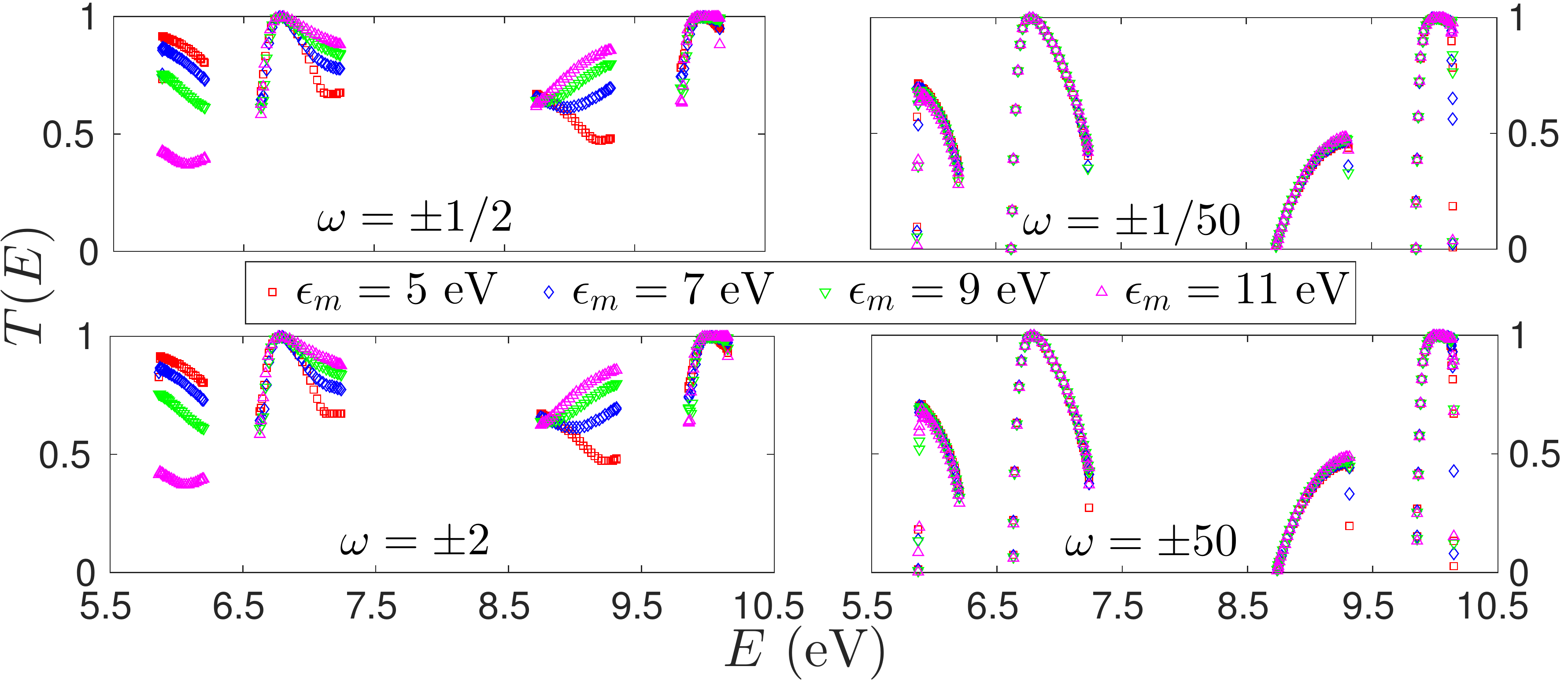}
\caption{Dependence of the upper envelope of the transmission coefficient on the energy center of the leads, $\epsilon_m$, for a system $u = 4$ sites per unit cell and $m = 30$ unit cells ($N = 120$), for symmetric coupling $\abs{\chi} =1$, in the strong (top left), very strong (top right), weak (bottom left) and very weak (bottom right) coupling regimes. The TB parameters of the system are   $(\epsilon_1,\epsilon_2, \epsilon_3,\epsilon_4) = (7, 9, 7.5, 8.5)$ eV and $(t_1,t_2,t_3,t_4) = (1.2, 0.9, 1, 0.8)$ eV. As the strength/weakness of coupling increases, the transmission profile becomes less dependent on the increase of $\epsilon_m$.}
\label{fig:TCWMomegaNOT1Em}
\end{figure}

\begin{figure} [h!]
\centering
\includegraphics[width=0.9\columnwidth]{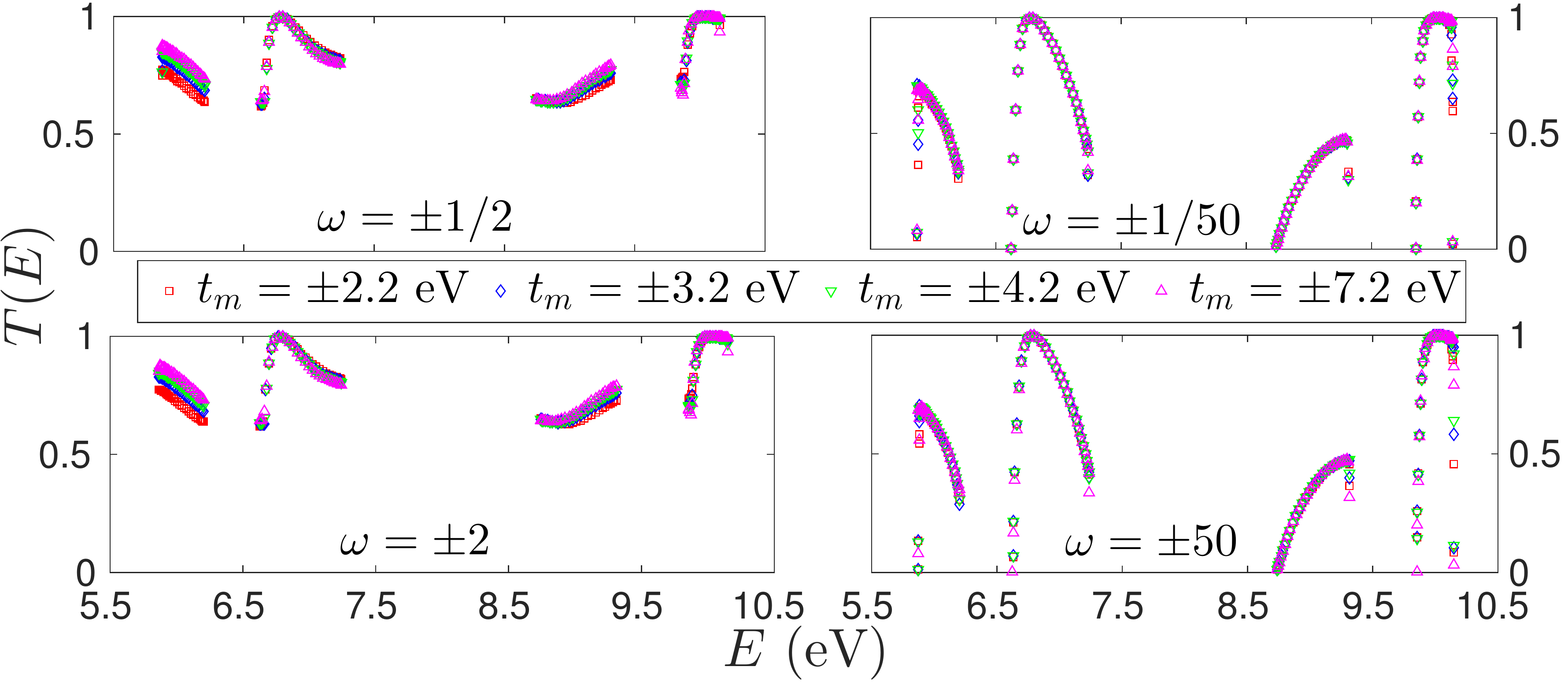}
\caption{Dependence of the upper envelope of the transmission coefficient on the bandwidth of the leads, as determined by $\abs{t_m}$, for a system $u = 4$ sites per unit cell and $m = 30$ unit cells ($N = 120$), for symmetric coupling $\abs{\chi} =1$, in the strong (top left), very strong (top right), weak (bottom left) and very weak (bottom right) coupling regimes. The TB parameters of the system are   $(\epsilon_1,\epsilon_2, \epsilon_3,\epsilon_4) = (7, 9, 7.5, 8.5)$ eV and $(t_1,t_2,t_3,t_4) = (1.2, 0.9, 1, 0.8)$ eV. As the strength/weakness of coupling increases, the transmission profile becomes less dependent on the increase of $\abs{t_m}$. Generally, increasing  $\abs{t_m}$ has less effect on the transmission profiles than increasing $\epsilon_m$.}
\label{fig:TCWMomegaNOT1tm}
\end{figure}

As it is obvious from the previous discussion, the term that incorporates the properties of the leads, $\cos(qa)$, is significant in Eq. \eqref{Eq:Lambdadef} only if the coupling asymmetry is of comparable magnitude with the coupling strength or weakness. In this case, the effect of the alteration of the leads' energy center and bandwidth to the transmission profile is not negligible. Both the upper and the lower envelope are affected. The lower envelope is affected in the same manner as for the ideal coupling condition (see Subsubsec. \ref{subsubsec:idealcouplingleads}). The upper envelope is also more affected by altering of $\epsilon_m$ that it is by altering $\abs{t_m}$. If the coupling asymmetry is much larger than its strength or weakness, the transmission profiles are remain almost unchanged. The above remarks are summarized in Figs. \ref{fig:TCWMomegaNOT1Em} and \ref{fig:TCWMomegaNOT1tm}, where the effect of increasing $\epsilon_m$ and $\abs{t_m}$, respectively, to the upper envelopes of the TC in the strong and weak coupling regime, is displayed, for the system of Fig. \ref{fig:UC4eigenvalues} with $m = 30$ unit cells ($N = 120$). 

\subsection{\label{subsec:OptimalCouplingCondition} Optimal Coupling Condition}
From the above discussion, we conclude that in the strong and weak coupling regimes, the transmission peaks sharpen and the TC does not generally reach the full transmission condition $T(E) = 1$. Hence, the ideal coupling condition $\abs{\omega} = 1$ is also the optimal coupling condition. This remark generalizes, for any periodic 1D lattice, and provides with a physical meaning, the optimal coupling condition reported in Ref. \cite{Macia:2005}, for a single stranded DNA segment with four sites per unit cell, in which all the hopping integrals of the DNA system were assumed to be equal.

\subsection{\label{subsec:TCuc1and2} The intrinsic interactions of the system}
The previous discussion for the effect of the coupling strength $\omega$ and the coupling asymmetry $\chi$ on the transmission profiles is specified in this Subsection for the two simplest cases of periodic 1D TB lattices ($u=1,2$). The effect of the intrinsic interactions of the system is also demonstrated.

\begin{figure*} [h!]
\centering
\includegraphics[width=\textwidth]{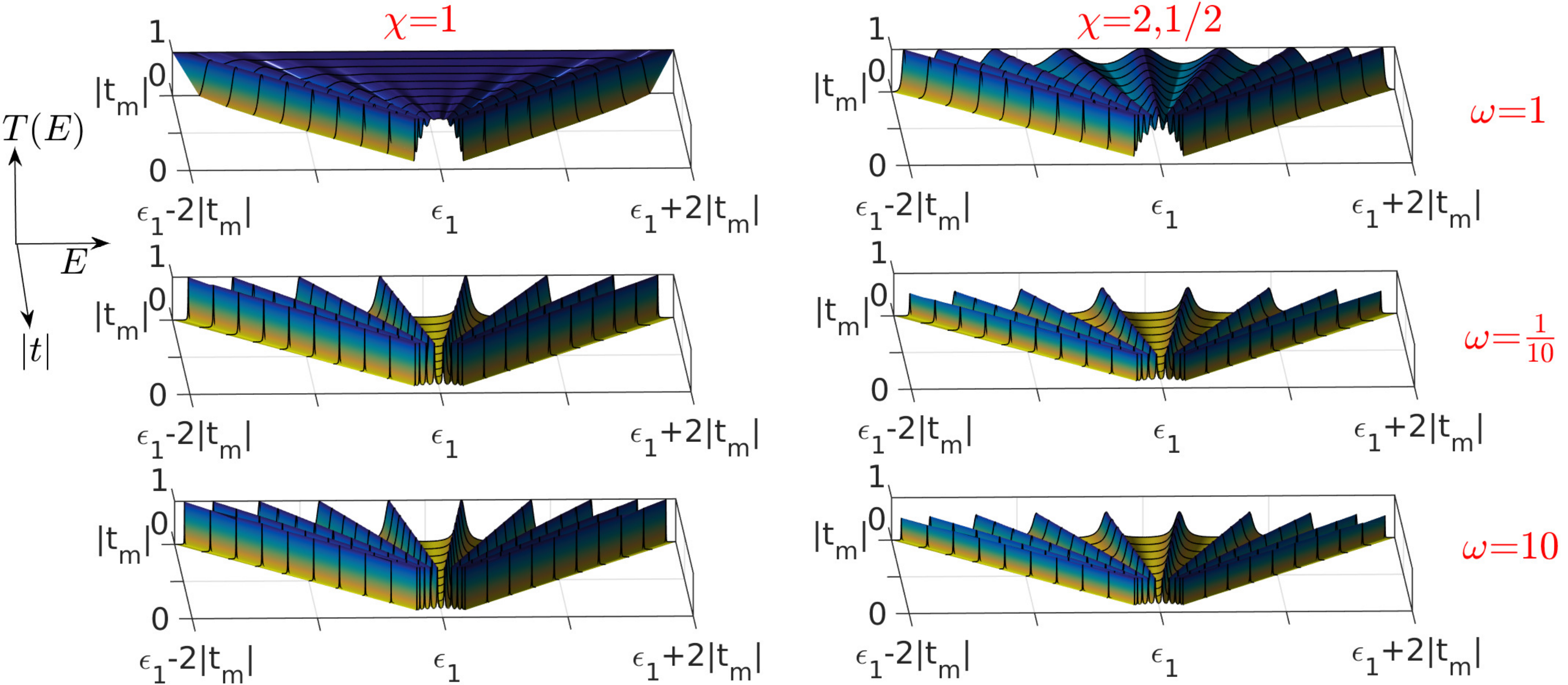}
\caption{Transmission coefficient of a periodic system with $u = 1$ site per unit cell and $m = N= 10$ unit cells, for ideal (top), strong (middle) and weak (bottom) coupling with the leads. (Left column) Symmetric coupling. (Right column) Asymmetric coupling. The on-site energy of the system coincides with that of the leads ($\epsilon_1 = \epsilon_m$) and the hopping integral of the system increases until it is equal to that of the leads.}
\label{fig:TCUC1}
\end{figure*}

In Fig. \ref{fig:TCUC1}, we present the TC of a system with $u=1$ site per unit cell and $m = N = 10$ unit cells, for the ideal, strong and weak coupling conditions, with or without coupling asymmetry, as the magnitude of the hopping integral of the system, $\abs{t_1}$, increases up to $\abs{t_m}$. We have chosen $\epsilon_1 = \epsilon_m$. The number of peaks is always $9$ for ideal coupling, $8$ for strong coupling and $10$ for weak coupling, as expected by the above discussion. It is also evident that the ideal and symmetric coupling conditions lead to the most efficient transmission, although in this simple case, the full transmission condition $T(E) =1$ is reached in all cases of symmetric coupling. In the upper left panel of Fig. \ref{fig:TCUC1} it can be seen that when $\abs{t_1} = \abs{t_m}$ the structures of the leads and the system become identical, hence the system is totally transparent.

\begin{figure*} [h!]
\centering
\includegraphics[width=\textwidth]{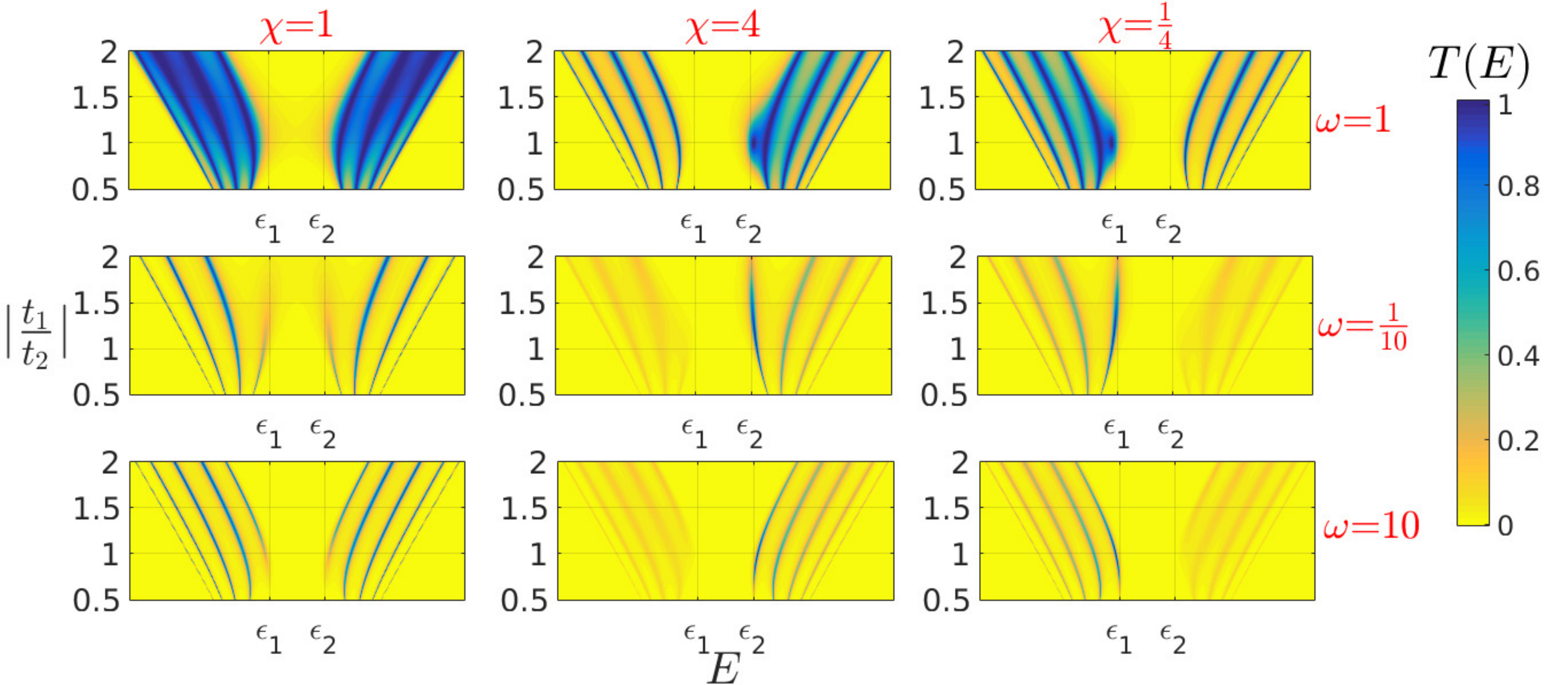}
\caption{Transmission coefficient of a periodic system with $u = 2$ sites per unit cell and $m = 5$ unit cells ($N = 10$), for ideal (top), strong (middle) and weak (bottom) coupling with the leads. (Left column) Symmetric coupling. (Middle column) Asymmetric coupling with $\abs{\chi} > 1$. (Right column) Asymmetric coupling with $\abs{\chi} < 1$. We have chosen $\epsilon_m = \frac{\epsilon_1+\epsilon_2}{2}$ and the bandwidth of the leads is chosen so as to contain all the eigenstates of the system.}
\label{fig:TCUC2}
\end{figure*}

In Fig. \ref{fig:TCUC2}, we present the TC of a system with $u=2$ sites per unit cell and $m = 5$ unit cells ($N = 10$), for the ideal, strong and weak coupling conditions, with or without coupling asymmetry, as the magnitude of the ratio between the hopping integrals of the system, $\abs{\frac{t_1}{t_2}}$, increases. We have chosen $\epsilon_m = \frac{\epsilon_1+\epsilon_2}{2}$ and the bandwidth of the leads, as determined by $\abs{t_m}$, is chosen so as to contain all the eigenstates of the system. Again, it is evident that the ideal and symmetric coupling condition leads to the most efficient transmission. For ideal and asymmetric coupling, except for the peaks of magnitude $1$ that occur in the zeros of $U_4(z)$, there is one additional peak near $E = \epsilon_1$ ($E = \epsilon_2$), when $\chi > 1$ ($\chi < 1$), i.e. in the region of the zeros of $M_2^{12}$ ($M_2^{21}$). This peak is of significant magnitude only when $\abs{t_1} \approx \abs{t_2}$. In the strong (weak) coupling regime, 8 (10) peaks occur, as expected; the peaks that are closer to the band gap vanish (emerge) as $\abs{\frac{t_1}{t2}}$ increases. When the coupling is asymmetric, transmission is enhanced only in one of the two bands.

\section{\label{sec:Conclusions} Conclusions}
We employed the transfer matrix method to obtain the energy eigenvalues of a periodic 1D TB model with a single orbital per site, $u$ sites per unit cell, and nearest neighbor interactions, for either cyclic or fixed boundaries. The solution of such a system is identical to the diagonalization of a real symmetric $u$-Toeplitz matrix of order $mu$, with or without perturbed upper left/lower right corners. The dispersion relation (cyclic boundaries) and the eigenspectrum (fixed boundaries) can be obtained with the help of the elements of UCTM and the Chebyshev polynomials of the second kind. The DOS was also obtained with the help of the UCTM. The properties of the eigenvalues and the DOS were discussed through representative examples for the simplest cases, i.e. for systems with $u = 1,2,3,4$, where the difference in the hopping integrals between different sites inside the unit cell was explicitly taken into account. Furthermore, we attached the periodic systems under examination to semi-infinite homogeneous leads, and determined the TC at zero bias. We showed that in general the ideal coupling condition is also the optimal coupling condition, and demonstrated the role of the coupling strength/weakness and asymmetry, as well as of the leads' properties, to the transmission profiles (number, position, sharpness of peaks). Finally, we demonstrated the effect of the intrinsic interactions of the system to the TC, for systems with $u=1,2$. 

\appendix
\begin{widetext}
\section{\label{ApA} Unit Cell Transfer Matrices for $u = 1,2,3,4$}
Here we provide the general form of the UCTM for systems with unit cell composed of $u = 1,2,3,4$ sites. Similar expressions can be produced for larger values of $u$, using Eqs. \eqref{Eq:UCTMrec}. For $u = 1$,
\begin{equation} \label{Eq:UCTM1}
M_1(E) = \begin{pmatrix} 
\frac{E-\epsilon_1}{t_1} & -1 \\
1 & 0
\end{pmatrix}.
\end{equation}
 For $u = 2$,
\begin{equation} \label{Eq:UCTM2}
M_2(E) = \begin{pmatrix} 
\frac{(E-\epsilon_2)(E-\epsilon_1)}{t_2t_1}-\frac{t_1}{t_2} & -\frac{E-\epsilon_2}{t_1} \\[6pt]
\frac{E-\epsilon_1}{t_1} & -\frac{t_2}{t_1}
\end{pmatrix}.
\end{equation}
For $u = 3$,
\begin{equation} \label{Eq:UCTM3}
M_3(E) = 
\begingroup
\renewcommand*{\arraystretch}{2}
\setlength\arraycolsep{10pt}
\begin{pmatrix} 
\frac{(E-\epsilon_3)(E-\epsilon_2)(E-\epsilon_1)}{t_3t_2t_1}-\frac{(E-\epsilon_3)t_1}{t_2t_3}-\frac{(E-\epsilon_1)t_2}{t_3t_1} & -\frac{(E-\epsilon_3)(E-\epsilon_2)}{t_1t_2} + \frac{t_2}{t_1}\\
\frac{(E-\epsilon_2)(E-\epsilon_1)}{t_1t_2}-\frac{t_1}{t_2} & -\frac{(E-\epsilon_2)t_3}{t_1t_2}
\end{pmatrix}.
\endgroup
\end{equation}
Finally, for $u = 4$,
\begin{equation} \label{Eq:UCTM4}
\begingroup
\renewcommand*{\arraystretch}{2}
\setlength\arraycolsep{20pt}
M_4 = \begin{pmatrix} 
\frac{(E-\epsilon_4)(E-\epsilon_3)(E-\epsilon_2)(E-\epsilon_1)}{t_4t_3t_2t_1}-\frac{(E-\epsilon_4)(E-\epsilon_3)t_1}{t_2t_3t_4} & -\frac{(E-\epsilon_4)(E-\epsilon_3)(E-\epsilon_2)}{t_3t_2t_1} + \frac{(E-\epsilon_4)t_2}{t_3t_1}  \\[3pt]
-\frac{(E-\epsilon_4)(E-\epsilon_1)t_2}{t_1t_3t_4}-\frac{(E-\epsilon_2)(E-\epsilon_1)t_3}{t_1t_2t_4}+\frac{t_1t_3}{t_2t_4} & +\frac{(E-\epsilon_2)t_3}{t_2t_1} \\[24pt]
\frac{(E-\epsilon_3)(E-\epsilon_2)(E-\epsilon_1)}{t_3t_2t_1}-\frac{(E-\epsilon_3)t_1}{t_3t_2}-\frac{(E-\epsilon_1)t_2}{t_1t_3} & -\frac{(E-\epsilon_3)(E-\epsilon_2)t_4}{t_3t_2t_1}+\frac{t_2t_4}{t_1t_3} 
\end{pmatrix}.
\endgroup
\end{equation}
\end{widetext}


\begin{thebibliography}{99}

\bibitem{SlaterKoster:1954}{J. C. Slater, and G. F. Koster, Simplified LCAO Method for the Periodic Potential Problem, Phys. Rev. \textbf{94} (1954) 1498.}

\bibitem{Harrison}{W. A. Harrison, Electronic Structure and the Properties of Solids: the Physics of the Chemical Bond, 2nd edition (Dover, New York, 1989)}

\bibitem{Papaconstantopoulos:2003}{D. A. Papaconstantopoulos, and M. J. Mehl, The Slater–Koster tight-binding method: a computationally efficient and accurate approach, J. Phys. Condens. Matter \textbf{15} (2003) R413.}

\bibitem{Roche-et-al:2003} {S. Roche, D. Bicout, E. Maci\'{a}, and E. Kats, Long Range Correlations in DNA: Scaling Properties and Charge Transfer Efficiency, Phys. Rev. Lett. \textbf{91} (2003) 228101.}

\bibitem{Roche:2003}{S. Roche, Sequence Dependent DNA-Mediated Conduction, Phys. Rev. Lett. \textbf{91} (2003) 108101.}

\bibitem{Shih:2008}{C. T. Shih, S. Roche, R. A. R\"{o}mer, Point-Mutation Effects on Charge-Transport Properties of the Tumor-Suppressor Gene p53, Phys. Rev. Lett. \textbf{100} (2008) 018105.}

\bibitem{Albuquerque:2014}{E. L. Albuquerque, U. L. Fulco, V. N. Freire, E. W. S. Caetano, M. L. Lyra, F. A. B. F. de Moura, DNA-based nanobiostructured devices: The role of quasiperiodicity and correlation effects, Physics Reports \textbf{535}, 139 (2014).}

\bibitem{Albuquerque:2005}{E. L. Albuquerque, M. S. Vasconcelos, M. L. Lyra, and F. A. B. F. de Moura, Nucleotide correlations and electronic transport of DNA sequences, Phys. Rev. E \textbf{71}, 021910 (2005).}

\bibitem{Berlin:2002}{Y. A. Berlin, A. L. Burin, M. A. Ratner, Elementary steps for charge transport in DNA: thermal activation vs. tunneling Chem. Phys. \textbf{275}, 61 (2002).}

\bibitem{Senthilkumar:2005}{K. Senthilkumar, F. C. Grozema, C. Fonseca Guerra, F. M. Bickelhaupt, F. D. Lewis, Y. A. Berlin, M. A. Ratner, and L. D. A. Siebbeles, Absolute Rates of Hole Transfer in DNA, J. Am. Chem. Soc. \textbf{127}, 14894 (2005).}

\bibitem{Cuniberti:2007}{G. Cuniberti, E. Maci\'{a}, A. Rodr\'{i}guez, and R. A. R\"{o}mer, Chapter: Tight-Binding Modeling of Charge Migration in DNA Devices, pp. 1-20 in T. Chakraborty (editor), Charge Migration in DNA Perspectives from Physics, Chemistry, and Biology, 2007, Springer-Verlag Berlin Heidelberg.}

\bibitem{Simserides:2014}{C. Simserides, A systematic study of electron or hole transfer along DNA dimers, trimers and polymers, Chem. Phys. \textbf{440}, 31 (2014).}

\bibitem{LKGS:2014}{K. Lambropoulos, K. Kaklamanis, G. Georgiadis, and C. Simserides, THz and above THz electron or hole oscillations in DNA dimers and trimers, Ann. Phys. (Berlin) \textbf{526}, 249 (2014).}

\bibitem{LChMKTS:2015} K. Lambropoulos, M. Chatzieleftheriou, A. Morphis, K. Kaklamanis, M. Theodorakou, and C. Simserides, Unbiased charge oscillations in B-DNA: Monomer polymers and dimer polymers, Phys. Rev. E \textbf{92}, 032725 (2015).

\bibitem{LKMTLGTCS:2016}{K. Lambropoulos, K. Kaklamanis, A. Morphis, M. Tassi, R. Lopp, G. Georgiadis, M. Theodorakou, M. Chatzieleftheriou, and C. Simserides, Wire and extended ladder model predict THz oscillations in DNA monomers, dimers and trimers, J. Phys. Condens. Matter \textbf{28}, 495101 (2016).}

\bibitem{LCMKLTTS:2016}{K. Lambropoulos, M. Chatzieleftheriou, A. Morphis, K. Kaklamanis, R. Lopp, M. Theodorakou, M. Tassi, and C. Simserides, Electronic structure and carrier transfer in B-DNA monomer polymers and dimer polymers: Stationary and time-dependent aspects of wire model vs. extended ladder model, Phys. Rev. E \textbf{94}, 062403 (2016).}

\bibitem{Nikerov:1976}{M. V. Nikerov, Electronic structure of carbyne, Journal of Structural Chemistry \textbf{17} (1976) 844.}

\bibitem{Tommasini:2007}{M. Tommasini, D. Fazzi, A. Milani, M. Del Zoppo, C. Castiglioni, and G. Zerbi, ffective hamiltonian for $\pi$ electrons in linear carbon chains, Chem. Phys. Lett. \textbf{450}, 86 (2007).}

\bibitem{Milani:2008}{A. Milani, M. Tommasini, and G. Zerbi, Carbynes phonons: A tight binding force field, J. Chem. Phys. \textbf{128}, 064501 (2008).}

\bibitem{Tommasini:2008}{M. Tommasini, A. Milani, D. Fazzi, M. Del Zoppo, C. Castiglioni, and G. Zerbi, Modeling phonons of carbon nanowires, Physica E \textbf{40}, 2570 (2008).}

\bibitem{La Magna:2009}{A. La Magna, I. Deretzis, and V. Privitera, Insulator-metal transition in biased finite polyyne systems, Eur. Phys. J. B \textbf{70}, 311 (2009).}

\bibitem{Dyachkov:2013}{P. N. D’yachkov, V. A. Zaluev, E. Yu. Kocherga, and N. R. Sadykov, Tight Binding Model of Quantum Conductance of Cumulenic and Polyynic Carbynes, J. Phys. Chem. C. \textbf{117}, 16306 (2013).}

\bibitem{Gover:1994}{M. J. C. Gover, The eigenproblem of a tridiagonal 2-Toeplitz matrix, Linear Algebra and its Applications \textbf{197-198} (1994) 63.}

\bibitem{Kouachi:2006}{S. Kouachi, Eigenvalues and eigenvectors of tridiagonal matrices, Electronic Journal of Linear Algebra \textbf{15} (2006) 115.}

\bibitem{Alvarez:2005}{R. \'Alvarez-Nodarse, J. Petronilho, N. R. Quintero, On some tridiagonal k-Toeplitz matrices: Algebraic and analytical aspects. Applications, J. Comput. Appl. Math. \textbf{184}, 518 (2005).}

\bibitem{Yueh:2008}{W.-C. Yueh, and S. S. Cheng, Explicit eigenvalues and inverses of tridiagonal Toeplitz matrices with four perturbed corners, ANZIAM J. \textbf{49}, 361 (2008).}

\bibitem{DaFonseca:2007}{C. M. da Fonseca, The characteristic polynomial of some perturbed tridiagonal k-Toeplitz matrices, Appl. Math. Sci. \textbf{1}, 59 (2007).}

\bibitem{Lee:1981}{D. H. Lee, and J. D. Joannopoulos, Simple scheme for surface-band calculations. I, Phys. Rev. B \textbf{23}, 4988 (1981).}

\bibitem{Stone:1981}{A. Douglas Stone, J. D. Joannopoulos, and D. J. Chadi,  Scaling studies of the resistance of the one-dimensional Anderson model with general disorder, Phys. Rev. B \textbf{24}, 5583 (1981).}

\bibitem{Molinari:1997}{L. Molinari, Transfer matrices and tridiagonal-block Hamiltonians with periodic and scattering boundary conditions, J. Phys. A: Math. Gen. \textbf{30}, 983 (1997).}

\bibitem{Molinari:2003}{L. Molinari, Spectral duality and distribution of exponents for transfer matrices of block-tridiagonal Hamiltonians, J. Phys. A: Math. Gen. \textbf{36}, 4081 (2003).}

\bibitem{Yeh:1977}{P. Yeh, A. Yariv, and C.-S. Hong, Electromagnetic 
propagation in periodic stratified media. I. General theory, J. Opt. Soc. Am. \textbf{67}, 423 (1977).}

\bibitem{Macia:2009}{E. Maci\'{a} Barber, Chapter 9, Mathematical Tools, in \textit{Aperiodic Structures in Condensed Matter, Fundamentals and Applications},  CRC Press, Taylor and Francis Group, LLC 2009, ISBN-13: 978-1-4200-6827-6.}

\bibitem{TsuandEsaki:1973}{R. Tsu, and L. Esaki, Tunneling in a finite superlattice, Appl. Phys. Lett. \textbf{22}, 562 (1973).}

\bibitem{Emberly:1998}{E. G. Emberly, and G. Kirczenow, Theoretical study of electrical conduction through a molecule connected to metallic nanocontacts, Phys. Rev. B \textbf{58}, 10911 (1998).}

\bibitem{Kostyrko:2002}{T. Kostyrko, An analytic approach to the conductance and I–V characteristics of polymeric chains, J. Phys. Condens. Matter \textbf{14}, 4393 (2002).}

\bibitem{Smithetal:2002}{D. R. Smith, S. Schultz, P. Marko\v s, and C. M. Soukoulis, Determination of effective permittivity and permeability of metamaterial from reflection and transmission coefficients, Phys. Rev. B \textbf{65}, 195104 (2002).}

\bibitem{Baringhausetal:2014}{J. Baringhaus, M. Ruan, F. Edler, A. Tejeda, M. Sicot, A. Taleb-Ibrahimi, A.-P. Li, Z. Jiang, E. H. Conrad, C. Berger, C. Tegenkamp, and W. A. de Heer, Exceptional ballistic transport in epitaxial graphene nanoribbons, Nature \textbf{506}, 349 (2014).}

\bibitem{Macia:2005}{E. Maci\'a, F. Triozon, and S. Roche, Contact-dependent effects and tunneling currents in DNA molecules,
Phys. Rev. B \textbf{71}, 113106 (2005).}

\bibitem{GuoXu:2007a}{A.-M. Guo and H. Xu, Effects of interbase electronic coupling and electrode on charge transport through short DNA molecules: A numerical study, Phys. Lett. A \textbf{364}, 48 (2007).}

\bibitem{Mason:2003}{J. C. Mason and D. C. Handscomb, \textit{Chebyshev Polynomials}, CRC Press, Taylor and Francis Group, LLC, 2003, ISBN: 0-8493-0355-9.}



\end{thebibliography}
\end{document}